\documentclass[journal,10pt]{IEEEtran}

\usepackage{ifpdf}
\usepackage{cite}
\usepackage{subfigure}
\ifCLASSINFOpdf
	\usepackage[pdftex]{graphicx}
\else
  	\usepackage[dvipdfmx]{graphicx, color}
\fi
\usepackage[cmex10]{amsmath}
\usepackage[caption=false,font=footnotesize]{subfig}

\ifCLASSOPTIONcaptionsoff
	\usepackage[nomarkers]{endfloat}
	\let\MYoriglatexcaption\caption
	\renewcommand{\caption}[2][\relax]{\MYoriglatexcaption[#2]{#2}}
\fi
\usepackage{url}
\usepackage{amssymb}
\usepackage{amsthm}
\usepackage{mathtools}
\usepackage{booktabs}
\usepackage{color}
\usepackage{bm,bbm}
\usepackage{algorithm,algpseudocode}
\algdef{SE}[DOWHILE]{Do}{DoWhile}{\algorithmicdo}[1]{\algorithmicwhile\ #1}
\algdef{SE}[SUBALG]{Indent}{EndIndent}{}{\algorithmicend\ }%
\algtext*{Indent}
\algtext*{EndIndent}

\usepackage{optidef}

\usepackage{comment}
\usepackage{multirow}

\newcommand{\taya}[1]{
	}
\newcommand{\yama}[1]{
	}

\theoremstyle{definition}

\title{ 
\fontsize{20}{28}\selectfont Distributed Heteromodal Split Learning \\ for Vision Aided mmWave Received Power Prediction
}

\author{
	Yusuke~Koda,~\IEEEmembership{Student~Member,~IEEE,}
	Jihong Park,~\IEEEmembership{Member,~IEEE,}
	Mehdi~Bennis,~\IEEEmembership{Senior Member,~IEEE}
	Koji~Yamamoto,~\IEEEmembership{Senior Member,~IEEE,}
	Takayuki~Nishio,~\IEEEmembership{Senior Member,~IEEE, and}
	Masahiro~Morikura,~\IEEEmembership{Member,~IEEE,}
	\thanks{
		Y. Koda, K. Yamamoto, T. Nishio, and M. Morikura are with the Graduate School of Informatics, Kyoto University, Yoshida-honmachi, Sakyo-ku, Kyoto 606-8501, Japan (e-mail: \{koda@imc.cce., kyamamot@, nishio@, nakashima@imc.cce.\}i.kyoto-u.ac.jp).
		J. Park is with the School of Information Technology, Deakin University, Geelong, VIC 3220, Australia (email: jihong.park@deakin.edu.au)
		M. Bennis is with the Centre for Wireless Communications, University of Oulu, 90014 Oulu, Finland and is also with the Department of Computer Science (e-mail: mehdi.bennis@oulu.fi).

		This work was supported in part by JSPS KAKENHI (Grant No. JP18H01442).
		This work was also supported in part by the Academy of Finland under Grant 294128, in part by the 6Genesis Flagship under Grant 318927, in part by the KvantumInstitute Strategic Project (SAFARI), in part by the Academy of Finland through the MISSION Project under Grant 319759, and in part by the NOKIA grant foundation.
	}
}

\begin{document}

\maketitle
\vspace{-4em}
\begin{abstract}
	The goal of this work is the accurate prediction of millimeter-wave received power leveraging both radio frequency (RF) signals and heterogeneous visual data from multiple distributed cameras, in a communication and energy-efficient manner while preserving data privacy.
	To this end, firstly focusing on data privacy, we propose heteromodal split learning with feature aggregation (HetSLAgg) that splits neural network (NN) models into camera-side and base station (BS)-side segments.
	The BS-side NN segment fuses RF signals and uploaded image features without collecting raw images.
	However, the usage of multiple visual data leads to an increase in NN input dimensions, which gives rise to 
	additional communication and energy costs.
	To overcome additional communication and energy costs due to image interpolation to blend different frame rates, we propose a novel BS-side manifold mixup technique that offloads the interpolation operations from cameras to a BS.
	Subsequently, we confront energy costs for operating a larger size of the BS-side NN segment due to concatenating image features across cameras and propose an energy-efficient aggregation method.
	This is done via a linear combination of image features instead of concatenating them, where the NN size is independent of the number of cameras.
	Comprehensive test-bed experiments with measured channels demonstrate that HetSLAgg reduces the prediction error by 44\% compared to a baseline leveraging only RF received power.
	Moreover, the experiments show that the designed HetSLAgg achieves over 20\% gains in terms of communication and energy cost reduction compared to several baseline designs within at most 1\% of accuracy loss.
\end{abstract}

\IEEEpeerreviewmaketitle

\begin{IEEEkeywords}
	Millimeter-wave communications, received power prediction, multi-modal deep learning, split learning, RGB-D image, beyond 5G.
\end{IEEEkeywords}

\section{Introduction} \label{sec:introduction}

\IEEEPARstart{P}{redicting} future millimeter-wave (mmWave) channels is crucial for enabling low-latency broadband communication in 5G and beyond\cite{xu2018revolution, hoyhtya2019database, park2020extreme, nishio_jsac}. 
However, this is a notoriously challenging problem due to the frequent and sudden mmWave channel state transitions from line-of-sight (LoS) to non-LoS, and vice versa\cite{maccartney2017flexible, pathloss}. 
Indeed, mobile blockages dictate these mmWave channel condition changes, yet mmWave radio frequency (RF) signals, e.g., received signal strength (RSS), hardly involve meaningful features related to blockage mobility. 
For this reason, predicting LoS-NLoS transitions struggles with insufficient features in the RF signal domain, making the accurate mmWave received power prediction a daunting task.

\begin{figure*}[t]
	\centering
	\includegraphics[width=0.85\textwidth]{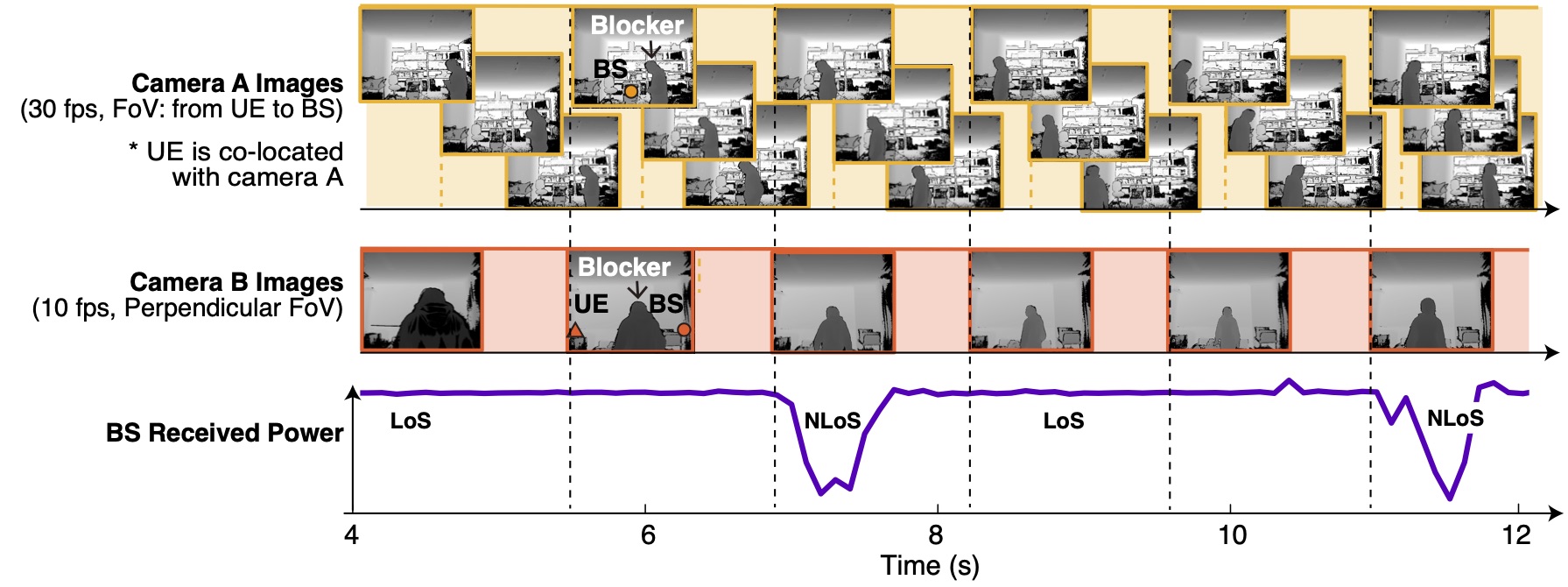}
	\caption{Illustrative example of visual data sequence from multiple distributed cameras and corresponding mmWave received power values.}
	\label{fig:illustrative_example}
	\vspace{-1.5em}
\end{figure*}

To complement such insufficient mmWave RF features, non-RF domain data can be utilized, such as location information\cite{zang2019managing, koda2018reinforcement}, motion sensory data\cite{bao2019motion} as well as visual data obtained from RGB-depth (RGB-D) cameras\cite{taha2017intelligent,nishio_jsac, koda2019handover}, which is the focus of our contribution.
As shown in Fig.~\ref{fig:illustrative_example}, a sequence of RGB-D image frames contain the object movement information within the camera's field of view (FoV), enabling accurate prediction of LoS-NLoS transitions. 
Leveraging this idea, we consider a base station (BS) communicating with an mmWave user equipment (UE) and  distributed RGB-D cameras having heterogeneous FoVs and frame rates, and aim to answer the following question: 
\textbf{how to acquire image data from multiple distributed cameras and fuse both RF and image modalities in a communication and energy efficient way for mmWave received power prediction while preserving privacy?}

Since deep learning is powerful in image feature extraction and fusion~\cite{goodfellow}, answering the aforementioned question boils down to developing communication and energy efficient distributed neural network (NN) architectures, training algorithms, and data pre/post-processing methods for mmWave received power prediction.
This is a complex task due to the use of multiple modalities and heterogeneous visual data, as we shall elaborate next.

\vspace{3pt}\noindent\textbf{1. Multimodal Fusion of RF and Image Data.}\quad
For a given LoS or NLoS condition, a sequence of RF signals contains more information on short-term received power fluctuations, whereas for LoS-NLoS transitions, a sequence of RGB-D images contains more information on the blockage mobility pattern. This mandates the need for utilizing both RF and image modalities. 
However, these multimodal data are not independent and identically distributed (IID), negating the effectiveness of ensembling-based distributed training methods~\cite{goodfellow} including federated learning~\cite{mcmahan2016communication}. 
To cope with non-IID multimodal data, we exploit the split learning (SL) framework~\cite{vepakomma2018split}, in which the features of both RF and image data samples are separately extracted, and then the resultant representations are transmitted and fused at the BS for the received power prediction. 
The feature extraction is performed at each camera via feeding forward convolutional and recurrent NN layers, while the representations (i.e., hidden layer activations) are fused at the BS using fully connected NN layers. 
Note that another benefit of SL is to preserve data privacy by exchanging hidden representations, without revealing raw images including privacy sensitive information, e.g., people's faces and travel records.

\vspace{3pt}\noindent\textbf{2. Heterogeneous FoVs and Frame Rates.}\quad
Heterogeneity of RGB-D camera images provides more useful features. 
As shown by Fig.~\ref{fig:illustrative_example}, multiple cameras' FoVs provide a wider range, whereas higher camera frame rates capture more detailed mobility patterns, all of which enhances the received power prediction accuracy. 
However, integrating heterogenous images induces the aforementioned problem of non-IID data distributions. 
To cope with heterogeneous FoVs, SL allows to fuse images from multiple cameras with different FoVs through their hidden representations, instead of the raw image samples. 
Furthermore, to address cameras' heterogeneous frame rates, we interpolate low-rate frames by linearly superposing two consecutive samples or their hidden representations, motivated by the Vanilla Mixup\cite{zhang2017mixup} and manifold Mixup data augmentation schemes\cite{verma2019manifold}, respectively.

\vspace{3pt}\noindent\textbf{3. Communication and Energy Efficient Interpolation.}\quad
RGB-D camera frame interpolation may improve the prediction accuracy, while increasing the size of each sequence for prediction. Processing larger sequences commonly requires a larger sized NN model consuming more energy \cite{han2015learning}, leading to a trade-off between energy efficiency and prediction accuracy. Furthermore, interpolating raw image samples (i.e., Vanilla Mixup) before transmission increases the communication payload sizes compared to interpolating hidden representations (i.e., manifold Mixup) after reception at the BS. Since Vanilla Mixup and manifold Mixup achieve different levels of prediction accuracy, it may yield a trade-off between communication efficiency and prediction accuracy. 
To address these trade-offs, we compare Vanilla and manifold Mixup methods as well as the case without interpolation (i.e., discarding higher-rate frames), in terms of communication, energy efficiencies as well as prediction accuracy. 

\vspace{3pt}\noindent\textbf{4. Energy Efficient Fusion.}\quad
Using principles of SL, the BS fuses the received representations from distributed cameras by concatenating them. 
The concatenation makes the BS NN layer size increase with the number of cameras, which is not scalable under limited energy. 
Alternatively, instead of the concatenation, we exploit the representation averaged across cameras so that the BS NN layer size becomes independent of the number of cameras. 
Such representation averaging is interpreted as inter-camera manifold Mixup, as opposed to frame rate interpolation using intra-camera manifold Mixup. 
We study the impacts and tradeoffs of inter-camera manifold Mixup and concatenation in terms of energy efficiency and prediction accuracy.

\subsection{Contributions}
The major contributions of this work are summarized as follows.
\begin{enumerate}
	\item 
	We propose a heteromodal split NN with feature aggregation (HetSLAgg) harnessing RF and heterogeneous visual data from distributed cameras for mmWave received power prediction while ensuring data privacy.
	The key idea for privacy preserving fusion of such heterogeneous inputs is to split the entire NN into camera-side and BS-side segments and to fuse the output image features from the camera-side NN segments into the BS along with RF features.
	This enables model training without sharing raw visual data, which ensures data privacy.
	Moreover, HetSLAgg harnesses RF and heterogeneous visual data because the entire NN structure is designed to calculate the predicted received power based on RF signal and multiple visual data inputs.

	\item 
	We develop a novel communication and energy efficient image interpolation to achieve a better tradeoff between prediction accuracy and communication/energy-efficiency while blending different frame rates.
	Therein, the key idea is to offload the interpolation operation from cameras to a BS based on Manifold Mixup, in which we interpolate the uploaded image features in the BS rather than the raw camera images.
	This avoids the increase in input dimensions in camera-side NN segments thereby avoiding additional energy-and-communication costs for calculating and uploading image features at cameras.

	\item 
	We develop an energy-efficient feature fusion procedure that prohibits an increase in the size of the BS-side NN segment when using more cameras while retaining meaningful heterogeneous image features.
	Therein, we take a linear combination of each component of the uploaded each image feature before feeding them into the BS-side NN segment rather than feeding the concatenated versions of the uploaded image features.
	This avoids the increase in the NN layer size in the BS and energy costs due to using more cameras.
	
	\item We demonstrate the effectiveness of the proposed methods via test-bed experiments with measured channels and RGB-D images. 
	The results show that: i) HetSLAgg benefits from RF and heterogeneous visual data by showing a lower root mean squared error (RMSE) by 20\% than the baseline of fusing RF and single visual data and by 33\% than the baseline of federated model averaging\cite{mcmahan2016communication};
	ii) With Manifold Mixup BS-side interpolation, both total communication and energy costs at cameras are reduced by over 20\% without any accuracy loss relative to Mixup camera-side interpolation baseline;
	iii) The proposed feature fusion method reduces the energy cost at the BS by 27\% relative to the baseline of concatenated feature fusion within 1\% accuracy loss.
\end{enumerate}

\subsection{Related Works and Organization}
\noindent\textbf{RF or non-RF Single Modality-based Wireless Systems.}\quad
For handover or positioning, RF-modalities, e.g., received power or channel state information, were studied\cite{mmwave_mdp,kaltiokallio2017three}.
For mmWave received power prediction, handover, prior studies in \cite{proactive3, nishio_jsac,koda2019handover, koda2020cooperative} leverage visual data to detect sudden LoS-and-NLoS transitions due to moving obstacles\cite{maccartney2017flexible, pathloss}.
While these aforementioned works demonstrate the feasibility of wireless systems benefitting from RF or non-RF modality, these works focus on the usage of a single modality.
Moreover, these works using visual data assume one single camera.
Unlike these works, we focus on the problem of fusing heterogeneous modalities, i.e., RF-signals and multiple visual data from distributed cameras.
Moreover, in contrast to these studies which do not take into account privacy in collecting visual data, we integrate RF and multiple visual data in a privacy-preserving manner.

\vspace{3pt}\noindent\textbf{RF and non-RF Modality Fusion.}\quad
Fusing RF and non-RF modalities, e.g., vision modalities, is proposed mainly in positioning of human pedestrians or robots to enhance prediction accuracy\cite{miyaki2007tracking, oskiper2010multi, alahi2015rgb, pham2016fusion}.
Therein, visual data is leveraged to achieve the best positioning accuracy while RF signals compensate for occluded cameras.
These works combine visual and RF modalities by simply taking weighted averages of the prediction results made with each modality \cite{miyaki2007tracking} or leveraging dual-stream convolutional neural networks (CNNs)\cite{pham2016fusion}.
Besides positioning, a seamless handover mechanism is proposed by fusing GPS information of mobile terminals and received signals from the mobile terminals\cite{ei2010trajectory}.
In addition to the fusion of RF and visual modalities, methods for fusing multiple streams of visual data are extensively studied in the literature to reconstruct three-dimensional objects and scenes\cite{aliakbarpour2016heterogeneous} or to compensate for single camera's limited FoV\cite{cheng2009multi, koda2020cooperative} or occlusions\cite{zhang2011segmentation}.
However, these studies do not consider the privacy in collecting highly private sensitive information, e.g., trajectory of humans viewed in visual data or mobile users tracked by GPS.
Unlike these studies, we study how to benefit from fusing RF-signals and multiple visual data in a privacy-preserving manner by leveraging a collaborative learning framework.

\vspace{3pt}\noindent\textbf{Federated Learning.}\quad
Collaborative learning frameworks exemplified  by FL\cite{mcmahan2016communication, park2019wireless, kairouz2019advances}, have recently attracted an increasing interest.
The key feature of collaborative learning is to train machine learning models in a distributed manner without sharing raw data samples, thereby preserving data privacy.
In classical FL, data owners hold and train ML models locally, and model parameters, e.g., the weight parameter of NN models, are averaged using a central server\cite{mcmahan2016communication} or other data owners\cite{elgabli2019gadmm}, after which these shared model parameters are integrated.
This prevents data owners from sharing raw data while enabling collaborative model training.. 
Several communication-efficient FL algorithms have been proposed such as NN pruning\cite{konevcny2016federated}, gradient compression\cite{lin2017deep, agarwal2018cpsgd}, and output distillation\cite{jeong2018communication}.
However, FL suffers from the challenging non-IID problem coming from the heterogeneity of data distribution across data owners.
Although the initial work regarding FL in \cite{mcmahan2016communication} alleviates the issue by increasing the number of local training iterations, this approach does not solve the issue when data distributions are totally different from one another. 
Note that this fact is experimentally verified in our work in the context of mmWave received power prediction leveraging multiple streams of visual data. 
In view of this, we aim at providing a more efficient collaborative learning framework leveraging the idea of split learning, that benefits from heterogeneous visual data from different cameras wherein the data distributions are totally different.
For a comprehensive survey and tutorial of federated learning and its applications, readers are encouraged to read \cite{park2019wireless, kairouz2019advances}.

\vspace{3pt}\noindent\textbf{Split Learning.}\quad
As another approach for collaborative learning, SL has been proposed in \cite{gupta2018distributed,vepakomma2018split}.
The key idea behind SL is to split the NN models on a per-layer basis and distribute a lower segment into data owners and an upper segment into another central entity.
Therein, the data owner does not need to exchange raw data but instead exchanges NN activations with the central entity during training.
In \cite{vepakomma2018split}, the authors conceptualized the fusion of multiple modalities with totally different data distributions, enabling privacy preserving model training.
However, this was a concept-level discussion, and the feasibility of applying SL based on multiple modalities to  wireless systems was not investigated.
Particularly, the design of communication and energy efficient SL remains an open problem, which is of importance because NN layer segments are interconnected wirelessly with a limited bandwidth, and some entities holding an NN segment are resource-limited, e.g., wireless cameras or mobile devices.
Our prior work in \cite{koda2020communication} proposes a communication-efficient SL framework for mmWave received power prediction fusing RF signals and single stream of visual data.
However, this prior work assumes a single camera, whereby the impact of using multiple cameras on prediction accuracy and communication and energy efficiency is not investigated.
To fill these voids, this work aims at designing an SL framework by fusing RF and visual data from multiple distributed cameras, while enabling model training in a communication and energy efficient manner.

\vspace{3pt}\noindent\textbf{Paper Organization.}\quad
The reminder of this paper is organized as follows.
In Section~\ref{sec:system_model}, we provide the proposed SL framework fusing RF and multiple visual modalities to enhance the accuracy of mmWave received power prediction.
In Section~\ref{sec:manifold_mixup}, we enhance the communication-and-energy efficiency of the proposed SL framework by incorporating the novel idea of interpolation of image features and aggregation of image features across distributed cameras.
In Section~\ref{sec:experiment}, we evaluate the proposed framework using experimentally obtained data set of multiple streams of depth images and RF received powers.
Finally, in Section~\ref{sec:conclusions}, we provide concluding remarks.
Note that some important notations are summarized in Table~\ref{table:notation} for sake of convenience to the reader.

\begin{table}
	\centering
	\caption{Summary of Notations}
	\label{table:notation}
	\begin{tabular}{ll}\toprule
		\textbf{Notation} & \textbf{Description}\\ \midrule
		\multicolumn{2}{l}{\emph{From Section~\ref{sec:system_model} to Section~\ref{sec:manifold_mixup}}:}  \\ 
		$\tau$ & Sampling period of camera~B\\
		$c$ & Ratio of frame rate of camera~A relative to camera~B \\
		$\bm{x}^{(i)}(t)$ & Image observed at time $t$ at camera~$i$\\
		$\bm{x}^{(i)}_k$ & Image observed at time $k\tau$ at camera~$i$ \\
		&	($\bm{x}^{(i)}_k\coloneqq \bm{x}^{(i)}(k\tau)$)\vspace{.2em}\\
		$\bm{a}^{(i)}_k$ & Image feature activation extracted from $\bm{x}^{(i)}_k$\\
		$P(t)$ & Received power at time $t$\\
		$P_k$	& Received power at time $k\tau$ ($P_k \coloneqq P(k\tau)$)\\
		$T_{\mathrm{back}}$ & Look-back time for feeding image sequence\\ 
		$T$ & Look-ahead time for received power\\
		$\hat{\bm{x}}^{(i)}_k$ &  Interpolated image at time $k\tau$ at camera~$i$\vspace{.2em}\\
		$\hat{\bm{a}}^{(i)}_k$ &  Interpolated feature activation at time $k\tau$ at camera~$i$\\
		$\lambda_k$ & Mixing coefficient for creating $\hat{\bm{x}}^{(i)}_k$ or $\hat{\bm{a}}^{(i)}_k$\\
		$\bm{a}^{(\mathrm{mix})}_k$ & Aggregated image feature activation \\
		&	from $\bm{a}^{(\mathrm{A})}_k$ and $\bm{a}^{(\mathrm{B})}_k$\\
		$\lambda_{\mathrm{agg}}$ & Mixing coefficient for creating $\bm{a}^{(\mathrm{mix})}_k$\\\midrule
		\multicolumn{2}{l}{\emph{Section~\ref{sec:experiment}}:}  \\ 
		$K$	& Number of image samples obtained in experiment\\
		$\mathcal{K}_{\mathrm{train}}, \mathcal{K}_{\mathrm{test}}$	&	Index set of training, test samples\\
		$R_{i, j}(t)$ & Transmission rate from node $i$ to $j$\\
		$P_{\mathrm{T}, i}$	&	Transmit power at node~$i$\\
		$P_{\mathrm{L}, i, j}$	&	Received power at node~$j$ from $i$ in LoS condition\\
		$A(t)$	&	Attenuation value of received power in NLoS condition\\
		& at time $t$\\
		$[k]$ &	Time interval $[(k - 1)\tau, k\tau]$\\
		$T_{\mathrm{FP}}[k]$  & Transmission latency for forward propagation signals\\
		&  within $[k]$\\
		$T_{\mathrm{BP}}[k]$  & Transmission latency for backward propagation signals\\
		&  within $[k]$\\
		$U(i)$	&	Indicator whether camera~$i$'s images are utilized\\	
		$D_{\mathrm{FP}}^{(i)}$  & Payload size of forward propagation signals\\
		&  from camera~$i$\\
		$D_{\mathrm{BP}}$  & Payload size of backward propagation signals\\
		$T_{\mathrm{comp}}$ & Total computation time for forward-backward\\
		&  propagation\\
		$T_{\mathrm{tot}}[k]$ & Total time required for forward-backward propagation\\
		& ($T_{\mathrm{tot}}[k] \coloneqq T_{\mathrm{FP}}[k] + T_{\mathrm{BP}}[k] + T_{\mathrm{comp}}$)\\
		$N[k]$	& Maximum number of forward-backward propagations\\
		& within $[k]$\\
		$T_n$	&	Elapsed time until the $n$th forward-backward \\
		& propagation is performed\\
		$N_{\mathrm{add}}, N_{\mathrm{mult}}$	&	Number of additions, multiplications\\
		$E_{\mathrm{add}}, E_{\mathrm{mult}}$	&	Energy per addition, multiplication \\
		$N_{\mathrm{param}}$	&	Number of parameters\\
		$E_{\mathrm{access}}$	&	Energy for loading one parameter\\\bottomrule
	\end{tabular}
\end{table}

\section{Split NN for Fusing RF and Heterogeneous Non-RF Modalities}
\label{sec:system_model}

\subsection{Proposed Heteromodal Split NN with Feature Aggregation}
\label{subsec:hetslagg}

\begin{figure}[]
	\centering
	\subfigure[Proposed: \textbf{HetSLAgg}.]{\includegraphics[width=\columnwidth]{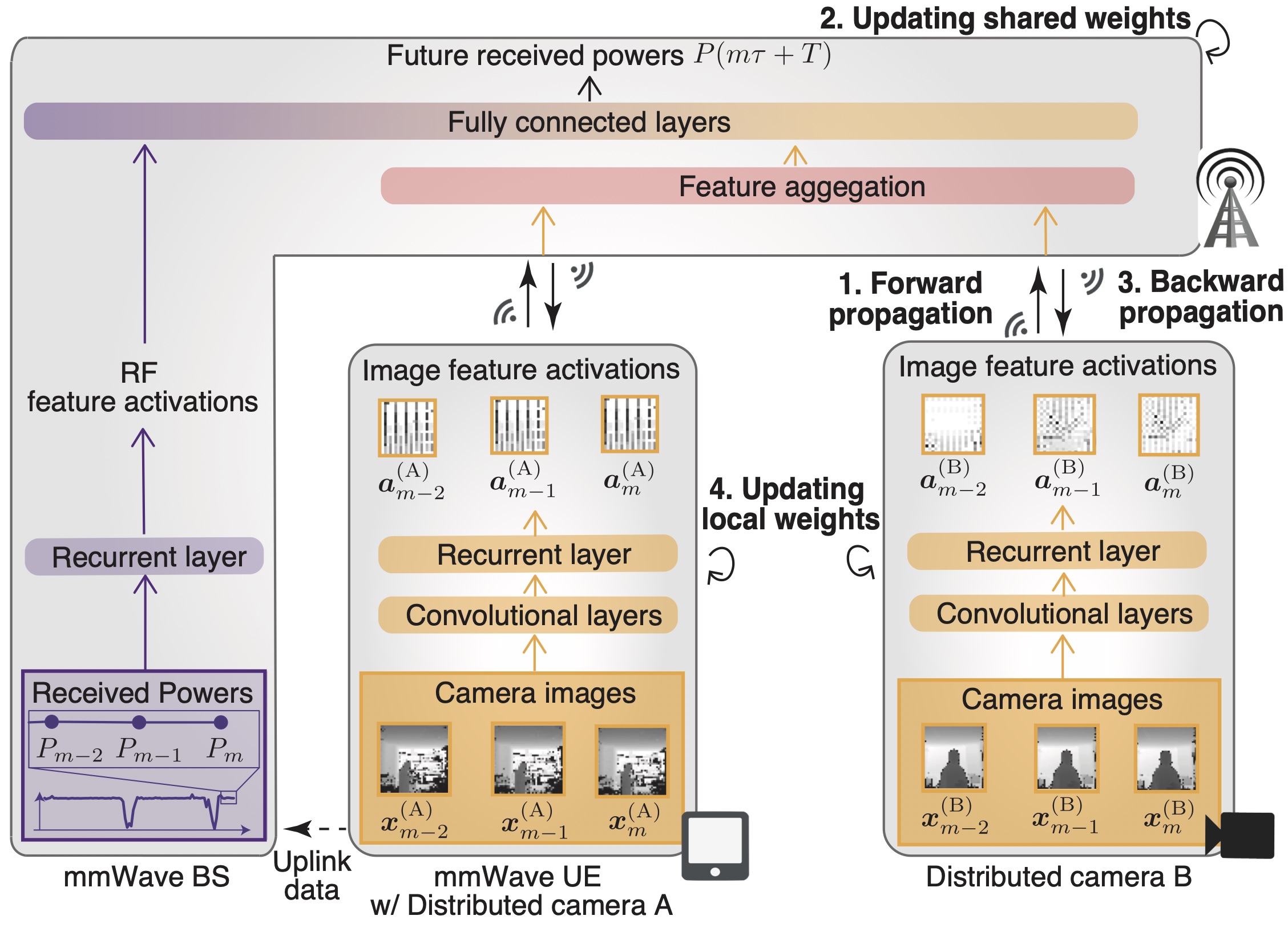}}\hspace{1em}
	\subfigure[Baseline:  HetSLFedAvg.]{\includegraphics[width=\columnwidth]{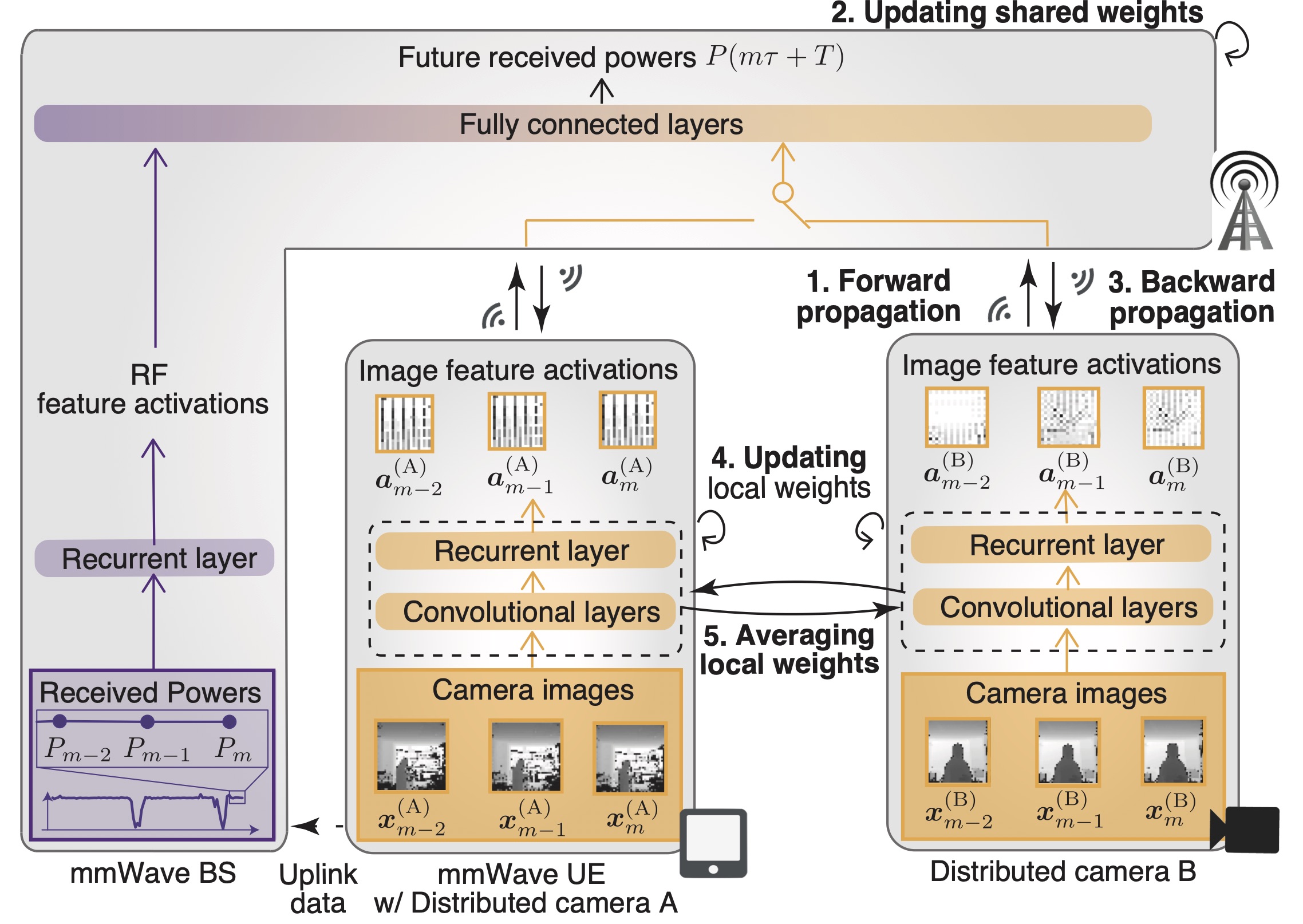}}
	\caption{
		Illustration of (a) the proposed hetero-modal split NN architecture with feature aggregation and (b) with federated model averaging.
		The architecture (b) is without feature aggregation across distributed cameras and is considered as a baseline to confirm the benefit of the feature aggregation in the proposed architecture (a).
	}
	\label{fig:system_model}
\end{figure}

We consider two distributed cameras termed camera~A and B without loss of generality as in Fig.~\ref{fig:system_model}.
Camera~A is embedded with an mmWave user equipment (UE) transmitting uplink data (e.g., uploading a large volume of data) to an mmWave BS, and we aim at predicting future received power values of the uplink signals.
Let $\bm{x}^{(i)}(t)$ and $P(t)$ denote the time-variant image obtained by camera~$i\in\{\mathrm{A}, \mathrm{B}\}$ and the uplink received power at the BS, respectively.
In this section, we consider that the sampling period of images or received powers are identical across the cameras and BS for the sake of simplicity, whereas this assumption is relaxed to different sampling periods in Section~\ref{sec:manifold_mixup}.
Each camera stores the image samples $\bigl(\bm{x}^{(i)}_{k}\bigr)_{k \in \{0, 1, \dots, m\}}$, where $\bm{x}^{(i)}_k\coloneqq \bm{x}^{(i)}(k\tau)$.
The terms $\tau$ and $m$ denote the sampling period of images and the index of the latest image sample, respectively.
Meanwhile, the BS stores received power samples $(P_{k})_{k \in \{0, 1, \dots, m\}}$, where $P_k\coloneqq P(k\tau)$.

In the proposed HetSLAgg, lower NN layer segments (e.g., convolutional and recurrent layers) are held in cameras storing visual data, and image feature activations, i.e., outputs of the camera-side NN layers, are aggregated in the BS.
Fig.~\ref{fig:system_model}(a) illustrates the proposed HetSLAgg.
To perform the prediction at each time $t = m\tau$, each camera uses the consecutive images within a look-back window $[m\tau - T_{\mathrm{back}}, m\tau]$, where $T_{\mathrm{back}}$ is the look-back time.
Specifically, each camera feeds images $(\bm{x}_k^{(i)})_{k\in \mathcal{K}'}$ into the first convolutional layer, where $\mathcal{K}'\coloneqq \{\,k\mid k\tau\in [m\tau - T_{\mathrm{back}}, m\tau], k\in \{0, 1, \dots, m\} \,\}$.
In the illustrative example in Fig.~\ref{fig:system_model}(a),  $T_{\mathrm{back}}$ is set as $2\tau$, where $\mathcal{K}' = \{m - 2, m - 1, m\}$, and three consecutive images $\bm{x}^{(i)}_{m - 2}, \bm{x}^{(i)}_{m - 1}, \bm{x}^{(i)}_m$ $\forall i \in\{\mathrm{A}, \mathrm{B}\}$ are leveraged.
Subsequently, each camera uploads image feature activations from the recurrent layer representing spatio-temporal features captured from each image, denoted by $(\bm{a}_k^{(i)})_{k\in \mathcal{K}'}$ that corresponds to $\bm{a}^{(i)}_{m - 2}, \bm{a}^{(i)}_{m - 1}, \bm{a}^{(i)}_m$ in Fig.~\ref{fig:system_model}(a).
The BS aggregates the image feature activations from both cameras and subsequently, feeds the aggregated activations into the fully connected layers.
At the same time, the BS feeds the received power samples $(P_k)_{k\in \mathcal{K}'}$, which corresponds to $P_{m - 2}, P_{m - 1}, P_m$ in Fig.~\ref{fig:system_model}(a), into the held recurrent layer.
The output of the recurrent layer, termed RF feature activations, is fed into the fully connected layers.
Finally, the fully connected layers output the predicted value of future received power $P(m\tau + T)$, where $T$ is the look-ahead time of the prediction.

The key features in training the entire NN model in HetSLAgg are forward and backward propagation over wireless channels.
The training procedure is also depicted in Fig.~\ref{fig:system_model}(a), which is based on the gradient descent optimization with back propagation commonly used to train NN models\cite{bishop}.
First, all cameras send their image feature activations to the BS over uplink wireless channels as illustrated in Fig.~\ref{fig:system_model}(a) with ``1.~Forward propagation''.
The BS updates the weight parameters of the held NN layers based on the gradients of them (``2.~Updating local weights'' in Fig.~\ref{fig:system_model}(a)).
Subsequently, the BS sends the gradients back to all cameras over downlink wireless channels, which is termed ``3.~Backward propagation'' in Fig.~\ref{fig:system_model}(a).
Based on the received gradients, the cameras complete ``4.~Updating local weights'', wherein they calculate the gradient of the weight parameters in the held NN layers and update the local weight parameters.

\subsection{Baseline: Heteromodal Split NN with Federated Model Averaging} 
\label{subsec:hetslfedavg_baseline}
The proposed HetSLAgg is compared with the baseline consisting of a split NN architecture without aggregating the image feature activations from distributed cameras.
This baseline architecture is illustrated in Fig.~\ref{fig:system_model}(b) and termed hetero-modal split NN with federated model averaging (HetSLFedAvg).
Being different from HetSLAgg, the BS does not perform the aggregation of the image feature activations uploaded from the cameras, feeding either activation into the fully connected layers along with the RF feature activation. 

The key feature in training HetSLFedAvg is that instead of image feature aggregation, HetSLFedAvg performs a model aggregation based on the federated model averaging\cite{mcmahan2016communication, li2018federated}, which is commonly used in FL.
First, one of the distributed cameras sends the output activation to the BS, and the BS calculates the gradients of the NN weights in the layers stored in the BS as illustrated in Fig.~\ref{fig:system_model}(b) with ``1.~Forward propagation''.
Based on the gradients, the BS updates the NN weights and sends the gradients back to the camera, which correspond to ``2.~Updating local weights'' and ``3.~Backward propagation'' in Fig.~\ref{fig:system_model}(b), respectively.
The cameras perform ``4.~Updating local weights'', wherein they calculate the gradient of the NN weights in the stored layers and complete the update of the weights.
This gradient descent optimization procedure is performed between the BS and all distributed cameras.
While performing this gradient optimization procedure, the distributed cameras perform ``5.~Averaging local weights'', wherein the weight parameters in the layers stored in each camera are exchanged and are averaged out across the cameras.

\section{HetSLAgg Design with Communication and Energy-Efficiency}
\label{sec:manifold_mixup}
In this section, we address two challenges of enabling communication and energy efficiency that comes from exploiting more than one distributed camera in the proposed HetSLAgg.
First, we address how to interpolate missing images in a communication and energy-efficient manner in different camera frame rate settings based on the key idea of interpolating the uploaded image feature activations at the BS.
This proposed BS-side interpolation technique is compared with camera-side raw image interpolation.
Second, we address the problem of how to feed the uploaded image feature activations into the fully connected layer in the BS while scaling to the number of distributed cameras in terms of energy costs at the BS.
Specifically, we develop an aggregation method that takes weighted averages of the uploaded features before feeding the image feature activations into the fully connected layer.
This proposed feature aggregation is compared with feeding a concatenated version of the uploaded image feature activations into the fully connected layer.

\begin{figure*}[]
	\centering
	\subfigure[Proposed: \textsf{MmixInt}.]{\includegraphics[width=0.32\textwidth]{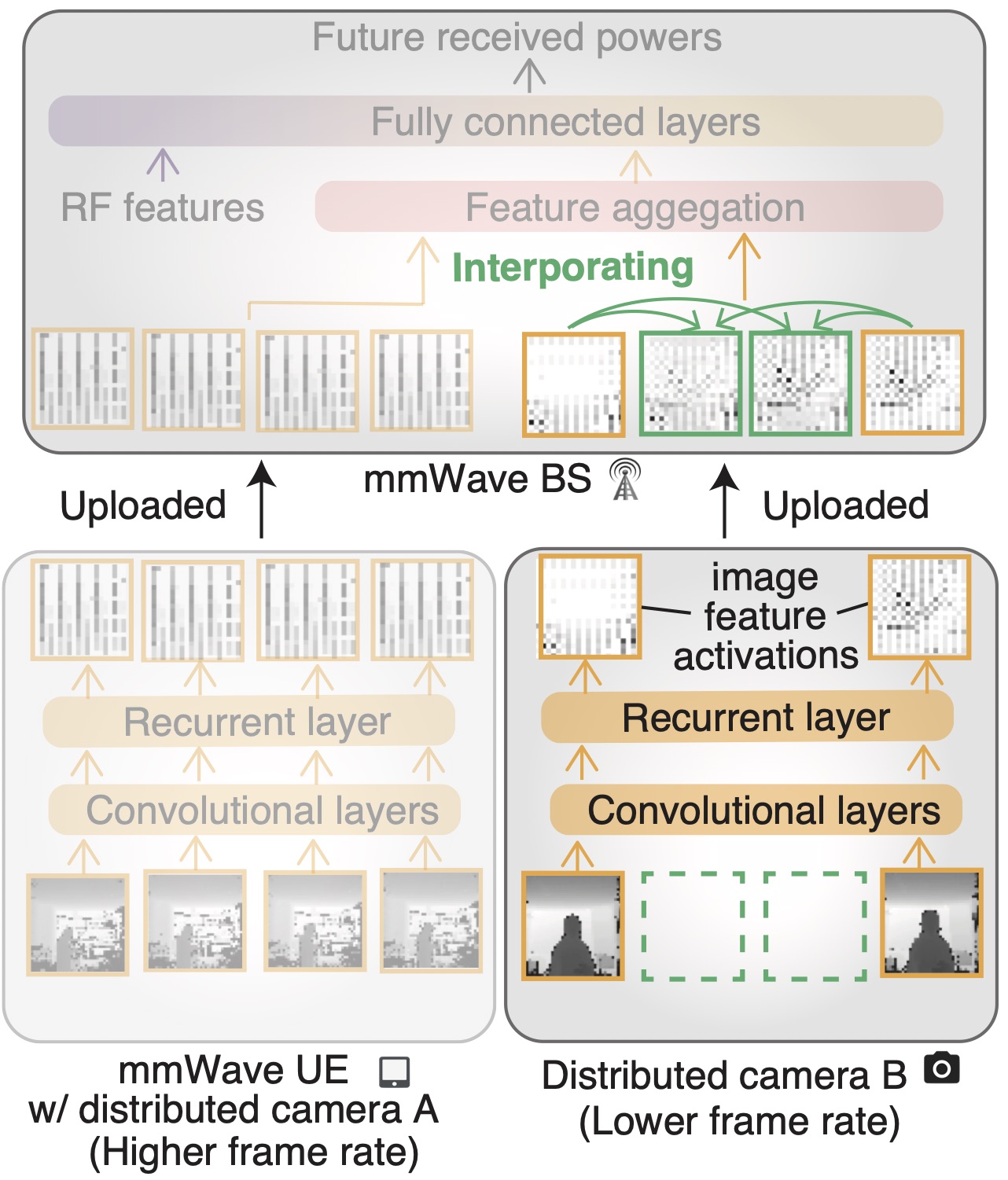}}\hspace{.5em}	
	\subfigure[Baseline~1: \textsf{MixInt}.]{\includegraphics[width=0.32\textwidth]{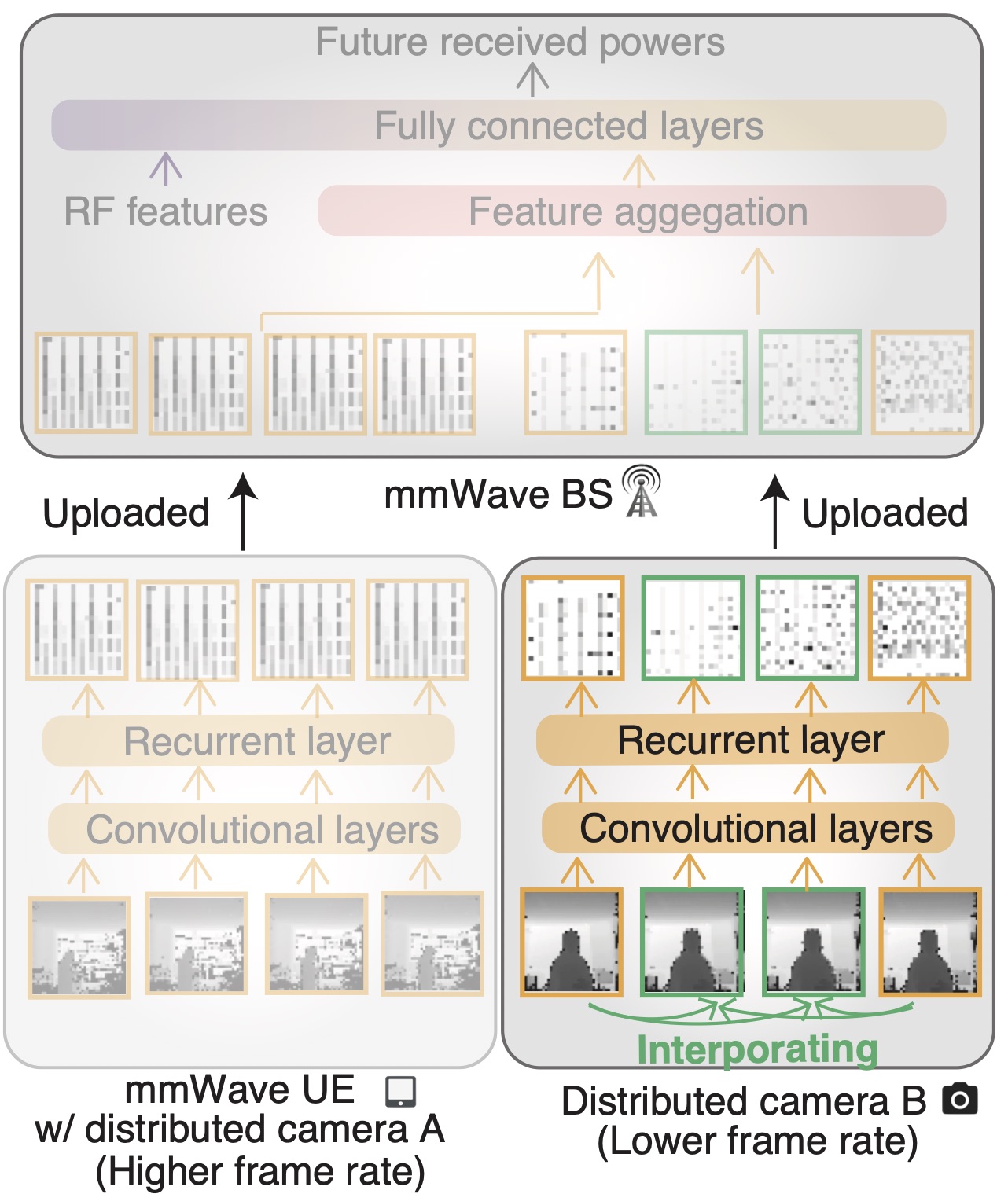}}\hspace{.5em}
	\subfigure[Baseline~2: \textsf{Disc}.]{\includegraphics[width=0.326\textwidth]{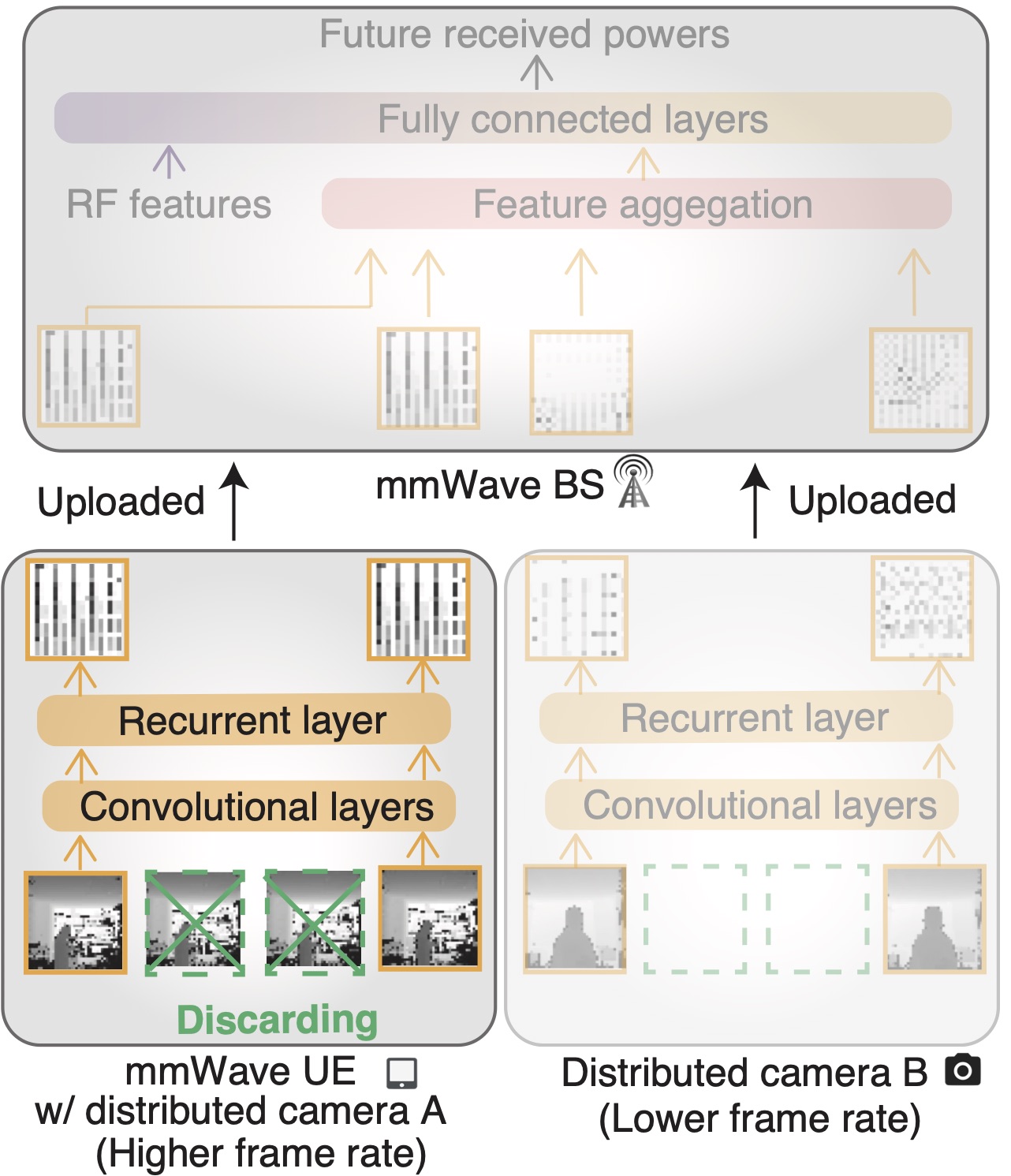}}
	\caption{Illustrative example of the proposed methods for balancing different frame rate based on BS-side manifold mixup-interpolation (\textsf{MmixInt}) when the ratio of frame rates $c = 3$. This is compared with two baselines based on camera-side mixup-based raw image interpolation (\textsf{MixInt}) and based on discarding images (\textsf{Disc}).}
	\label{fig:interpolation}
\end{figure*}

\subsection{Communication and Energy-Efficient Feature Interpolation for Balancing Frame Rate Difference with Manifold Mixup}
\label{subsec:interpolation}
Without loss of generality, we consider the case where two distributed cameras, termed camera~A and B, are exploited, and camera~A has a $c$ times higher frame rate than camera~B, where $c\in\mathbb{R}$.
Let $\tau$  denote the sampling period of camera~B, and $\bm{x}^{(i)}_{k}\coloneqq \bm{x}^{(i)}(k\tau)$ for $i\in\{\mathrm{A}, \mathrm{B}\}$.
Considering that each camera stores images in the time duration of $[0, m\tau]$, camera~B and camera~A store the image samples $\bigl(\bm{x}^{(\mathrm{B})}_{k}\bigr)_{k\in \mathcal{K}_{\mathrm{B}}}$ and $\bigl(\bm{x}^{(\mathrm{A})}_{k}\bigr)_{k \in \mathcal{K}_{\mathrm{A}}}$, respectively, where $\mathcal{K}_{\mathrm{B}}\coloneqq \{0, 1, 2, \dots, m\}$ and $\mathcal{K}_{\mathrm{A}}\coloneqq \{0, 1/c, 2/c, \dots, \lfloor mc \rfloor/c\}$.
The function $\lfloor\cdot\rfloor$ denotes the floor function.
As in Section~\ref{sec:system_model}, to perform the prediction at $t = m\tau$, we utilize the images within the look-back window $[m\tau - T_{\mathrm{back}}, m\tau]$.
Therein, the images $\bigl(\bm{x}^{(i)}_{k}\bigr)_{k\in \mathcal{K}'_{i}}$ for $i\in\{\mathrm{A}, \mathrm{B}\}$ are input for the prediction, where $\mathcal{K}'_{i}\coloneqq \{\,k\mid k\tau\in [m\tau - T_{\mathrm{back}}, m\tau], k\in \mathcal{K}_i\,\}$.
Note that $\bigl(\bm{x}^{(\mathrm{B})}_{k}\bigr)_{k\in \mathcal{K}'_{\mathrm{B}}}$ contains fewer images than  $\bigl(\bm{x}^{(\mathrm{A})}_{k}\bigr)_{k\in \mathcal{K}'_{\mathrm{A}}}$ owing to the lower frame rate as depicted in Fig.~\ref{fig:interpolation}.
In the following discussion, we interpolate the missing images in camera~B, i.e., $\bigl(\bm{x}^{(\mathrm{B})}_{k}\bigr)_{k\in \mathcal{K}'_{\mathrm{A}}\setminus\mathcal{K}'_{\mathrm{B}}}$.

The key idea behind the proposed framework is to interpolate the image feature at the BS with a linear combination of uploaded image feature activations using the idea of Manifold Mixup\cite{verma2019manifold}.
In the proposed Manifold Mixup-based image interpolation, rather than interpolating raw images in camera~B, the missing image feature activations are interpolated at the BS to reduce both the upload payload size and energy cost required for the procedure in convolutional layers as illustrated in Fig.~\ref{fig:interpolation}(a),.
Let the output activation corresponding to image $\bm{x}^{(i)}_k$ be denoted by $\bm{a}^{(i)}_k$.
Each missing output activation $\bm{a}^{(\mathrm{B})}_k$ for $k\in \mathcal{K}'_{\mathrm{A}}\setminus \mathcal{K}'_{\mathrm{B}}$ is interpolated by:
\begin{align}
	\label{eq:manifold_mixup}
	\hat{\bm{a}}^{(\mathrm{B})}_k = \lambda_k \bm{a}^{(\mathrm{B})}_{k'} + (1 - \lambda_k) \bm{a}^{(\mathrm{B})}_{k' + 1}, 
\end{align}
where $\hat{\bm{a}}^{(\mathrm{B})}_k$ is the interpolated output activation, and $k'\coloneqq \max\{\,k''\mid k'' \in \mathcal{K}'_{\mathrm{B}}, k'' \leq k\,\}$.
The term $\lambda_k\in (0, 1)$ is the interpolation factor associated with the index $k$.
In the experiment discussed in the Section~\ref{sec:experiment}, we set $\lambda_k$ as $k' + 1 - k$, which corresponds to piecewise linear interpolation and show that even this lightweight interpolation performs better than the following baselines.
For the sake of simplicity, we term the proposed interpolation as \textsf{MmixInt}, hereinafter.

Proposed \textsf{MmixInt} is compared with the following two baseline frameworks: camera-side raw image interpolation with Vanilla Mixup termed \textsf{MixInt} and camera-side image discarding termed \textsf{Disc}.
The first framework interpolates the missing original images in camera~B as depicted in Fig.~\ref{fig:interpolation}(b) with a linear combination of the obtained original images borrowing the idea of Mixup\cite{zhang2017mixup}. 
Therein, each missing image $\bm{x}^{(\mathrm{B})}_k$ for $k\in \mathcal{K}'_{\mathrm{A}}\setminus \mathcal{K}'_{\mathrm{B}}$ is interpolated by:
\begin{align}
	\label{eq:mixup}
	\hat{\bm{x}}^{(\mathrm{B})}_k = \lambda_k \bm{x}^{(\mathrm{B})}_{k'} + (1 - \lambda_k) \bm{x}^{(\mathrm{B})}_{k' + 1},
\end{align}
where $\hat{\bm{x}}^{(\mathrm{B})}_k$ is the interpolated images, and the other variants are consistent with \eqref{eq:manifold_mixup}.
The second framework is to discard the images $\bigl(\bm{x}^{(\mathrm{A})}_{k}\bigr)_{k\in \mathcal{K}'_{\mathrm{A}}\setminus\mathcal{K}'_{\mathrm{B}}}$ as depicted in Fig.~\ref{fig:interpolation}(c), wherein the images $\bigl(\bm{x}^{(\mathrm{A})}_{k}, \bm{x}^{(\mathrm{B})}_{k}\bigr)_{k\in \mathcal{K}'_{\mathrm{B}}}$ are utilized for the prediction.

\subsection{Energy-Scalable BS-Side Feature Aggregation with Manifold Mixup}
\label{subsec:aggregation}
\begin{figure*}[]
	\centering
	\subfigure[Proposed: \textsf{MmixAgg}.]{\includegraphics[width=0.42\textwidth]{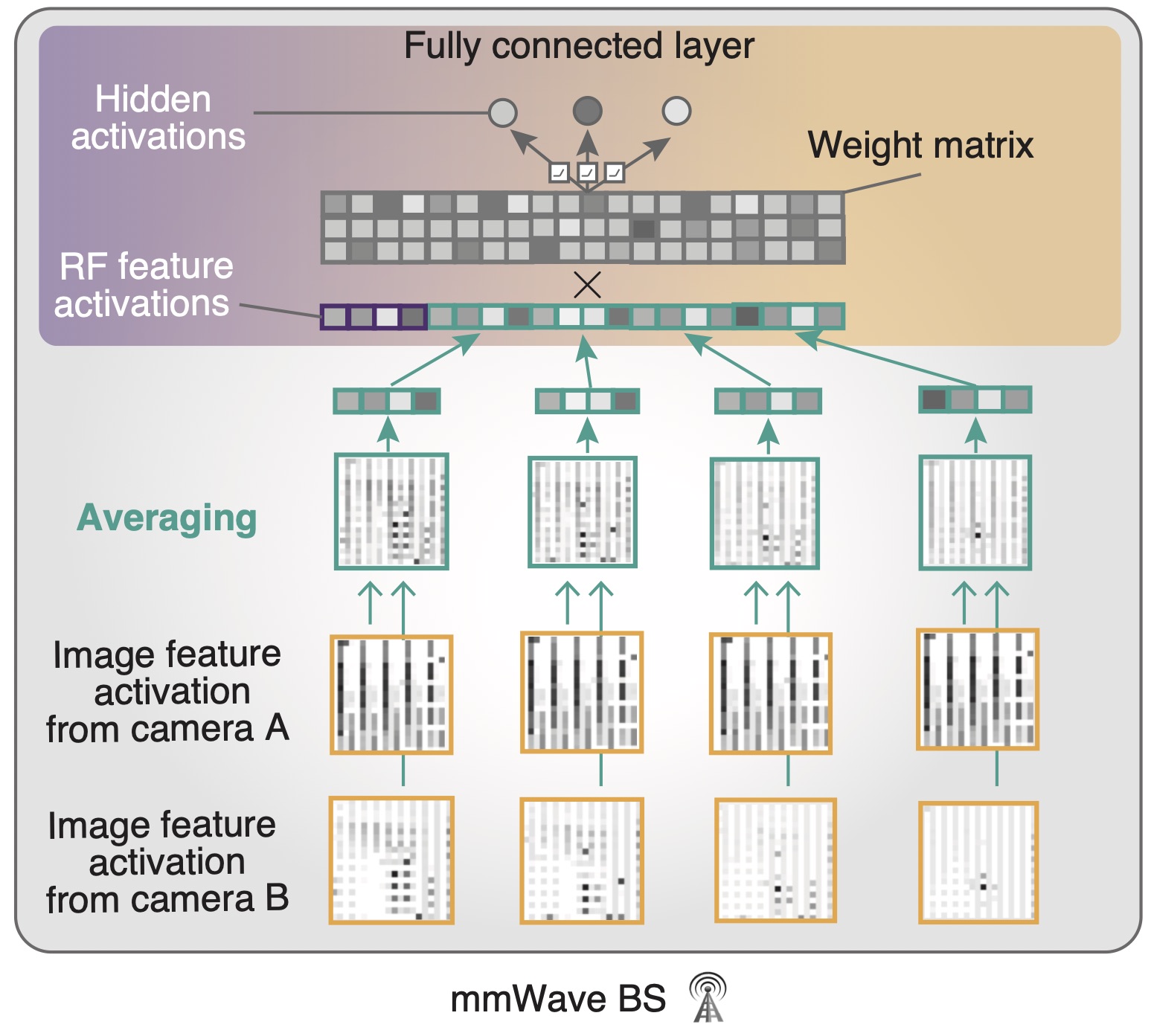}}\hspace{2em}
	\subfigure[Baseline: \textsf{ConcAgg}.]{\includegraphics[width=0.45\textwidth]{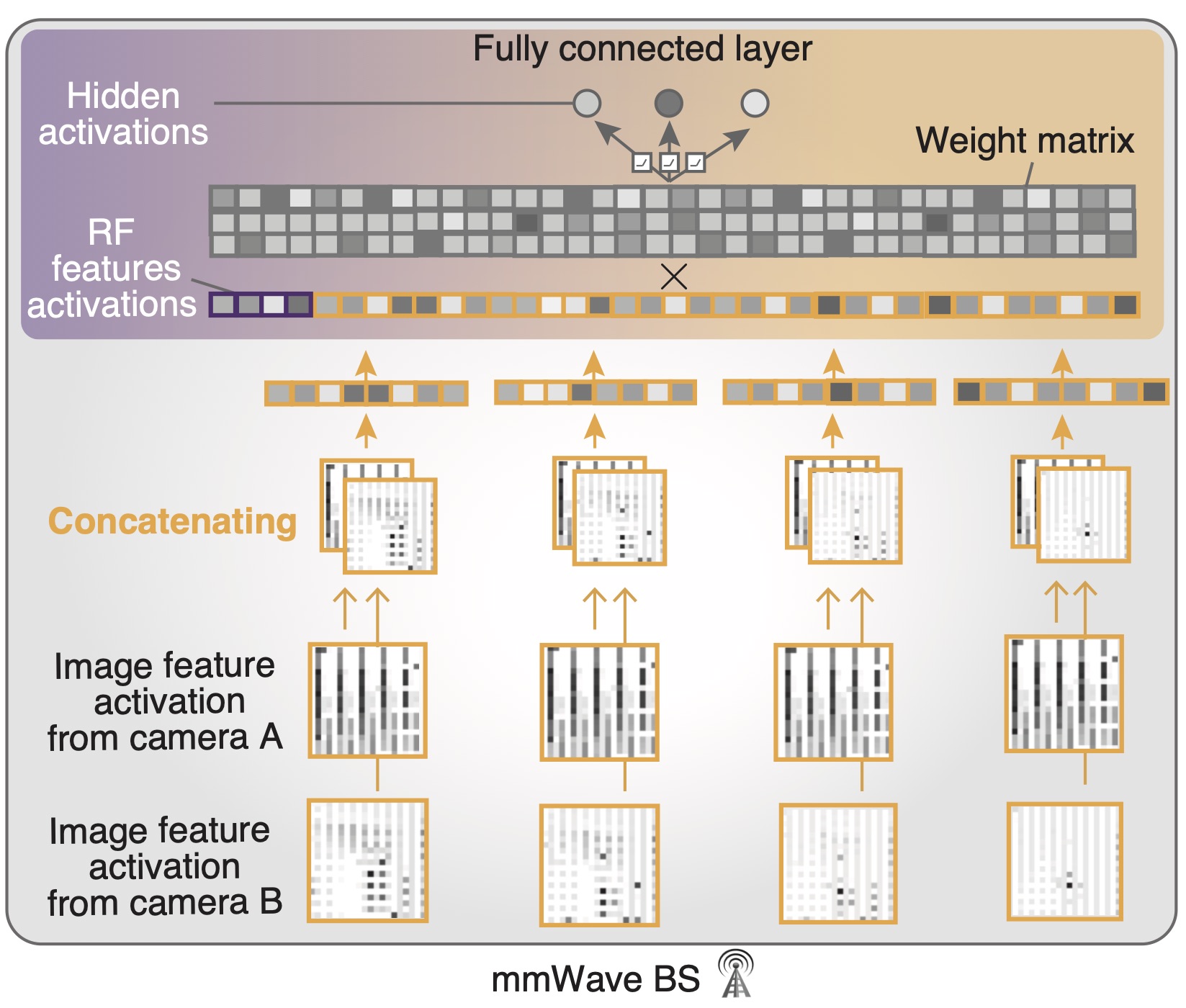}}
	\caption{Proposed method for aggregating image feature activations based on manifold mixup (\textsf{MmixAgg}) and baseline method based on concatenation (\textsf{ConcAgg}).}
	\label{fig:aggregation_method}
\end{figure*}
To realize the energy-scalability due to the increase in the number of cameras, we design a method for aggregating the uploaded image feature activations so that the input dimension of the fully connected layer in the BS does not scale up according to the number of cameras.
Therein, the number of weight parameters of the fully connected layer does not scale up, resulting in lower power consumptions.
As in the previous section, we consider two distributed cameras without loss of generality.

In the proposed feature aggregation method, we take a weighted average of the uploaded image feature activations as depicted in Fig.~\ref{fig:aggregation_method}(a).
Let $\bm{a}^{(\mathrm{mix})}_k$ denote the vectorized form of the aggregated feature activations which is fed into the fully connected layer while let $\bm{a}^{(\mathrm{A})}_k$ and $\bm{a}^{(\mathrm{B})}_k$ denote the vectorized form of the image feature activations uploaded form cameras~A and B, respectively.
Formally, in the proposed aggregation method, the aggregated image feature activation is given by:
\begin{align}
	\bm{a}^{(\mathrm{mix})}_k = \lambda_{\mathrm{agg}} \bm{a}^{(\mathrm{A})}_k + (1 - \lambda_{\mathrm{agg}})\bm{a}^{(\mathrm{B})}_k, 
\end{align}
where $\lambda_{\mathrm{agg}}$ denotes the weight of the averaging and is set as $1/2$ in the experiment in Section~\ref{sec:experiment}.
It should be noted that such feature averaging is interpreted as inter-camera Manifold Mixup, and hence, we term the proposed aggregation method as \textsf{MmixAgg} hereinafter.

The proposed \textsf{MmixAgg} is compared with the baseline of concatenating the uploaded image feature activations.
This baseline feeds the full version of the uploaded image feature activations into the fully connected layer in the BS as illustrated in Fig.~\ref{fig:aggregation_method}(b), and hence, the input dimension of the fully connected layer scales up according to the number of distributed cameras.
The experimental evaluation demonstrates that despite retaining the full version of the uploaded image feature activations, this baseline does not necessarily exhibit better prediction performance than the proposed manifold mixup-based aggregation method, which is discussed in Section~\ref{sec:experiment}.
We term this baseline method as \textsf{ConcAgg} hereinafter.

\section{Experimental Evaluations}
\label{sec:experiment}

\subsection{Experimental Setup}
\begin{figure}[t]
	\centering
	\includegraphics[width=\columnwidth]{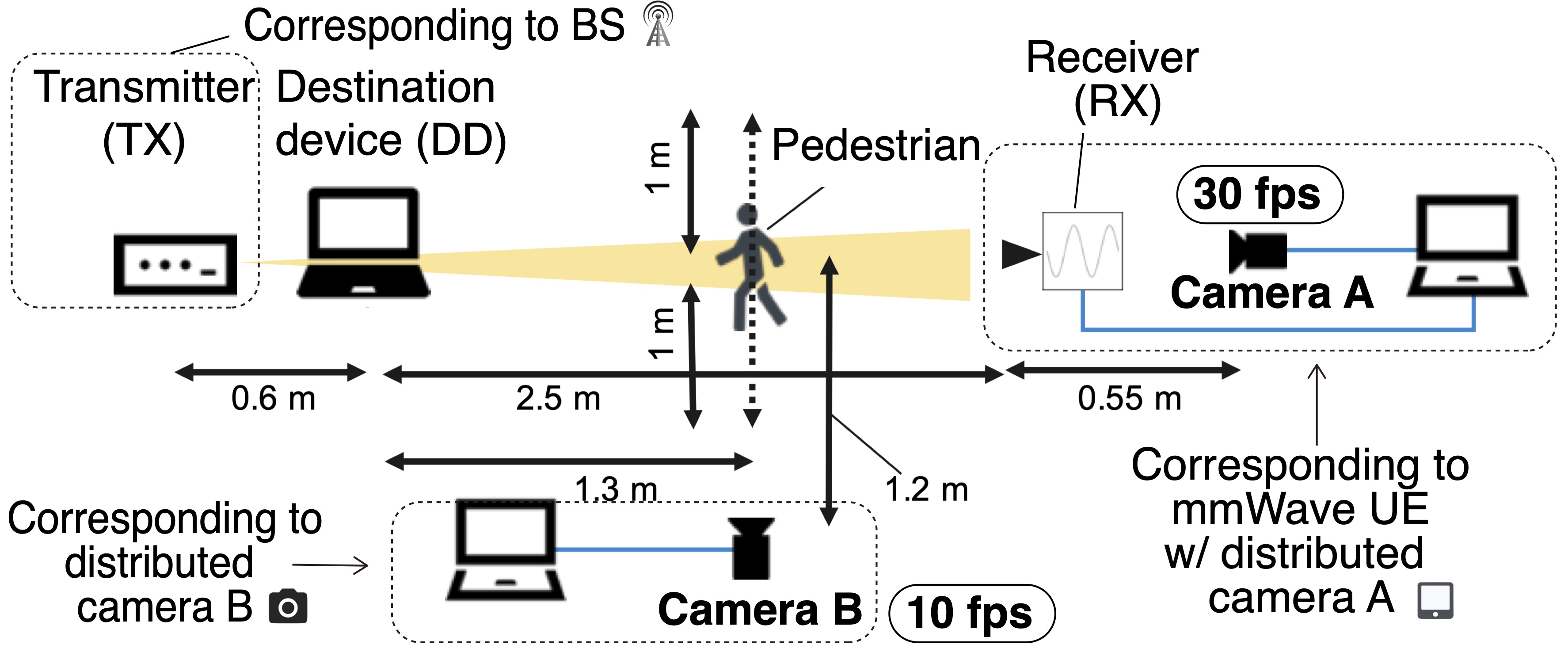}
	\caption{Experimental environment for measuring communication channel and depth images from two different cameras termed camera~A and B having different frame rates. 
	The communication channel is intermittently blocked by one pedestrian.
	Camera~A and B correspond to distributed camera~A and B in Fig~\ref{fig:system_model}, respectively.}
	\label{fig:experimental_environment}
\end{figure}

\noindent\textbf{Datasets.}\quad 
The training and evaluation is performed using a data set of received powers and depth images obtained in a real-world experiment.
The experimental environment is shown in Fig.~\ref{fig:experimental_environment}.
We deployed a transmitter (TX), the destination device (DD) for the frame transmission of the TX, receiver (RX), and two cameras termed camera~A and camera~B.
As the TX and DD, we utilized commercial products of an IEEE 802.11ad access point and station, respectively.
As the RX, we utilized the measurement device developed in \cite{koda_measurement2}.
The RX is equipped with a horn antenna with directivity gain of 24\,dBi and the half-power beam width (HPBW) of 11 degree while the TX is equipped with an array antenna with size of 16, directivity gain of approximately 8\,dBi, and HPBW of approximately 15\,degree.
As both camera~A and B, we utilized Kinect sensors\cite{kinect2}, which is capable of obtaining depth images with the resolution of $512\times 480$.
Camera~A is deployed behind the DD while camera~B is deployed apart from the TX and RX, wherein the viewing angles of camera~A and camera~B are orthogonal with each other.
The TX and RX correspond to the BS and UE in Fig.~\ref{fig:experimental_environment}, respectively, and camera~A and B correspond to the distributed camera~A embedded into the UE and distributed camera~B, respectively.

We conduct the measurement as in \cite{koda_measurement2} and obtain a time series of received powers.
The TX transmits signals at the carrier frequency of 60.48\,GHz towards the DD, and subsequently, the RX behind the DD receives the signals and measures the power of the signals.
While the signal transmission, one pedestrian walks across the path between the DD and RX and intermittently blocks the LoS path between them.
The purpose of this arrangement is to prevent the beam tracking of the TX, which is discussed in detail later.
In the RX, the time-variance of the mmWave received powers due to a moving pedestrian is measured.
While the measurement, the two cameras obtain depth images viewing the pedestrian from different angles with different time resolutions.
The frame rate of camera~A is 30\,frame per second (fps) while that of camera~B is 10\,fps. 

It should be noted that we examine HetSLAgg where only a single pedestrian causes blockage events for the following two reasons.
Firstly, the objective of this experiment is to demonstrate that HetSLAgg benefits from RF received powers and multiple visual data under blockage events, wherein it is sufficient to examine HetSLAgg where a single pedestrian causes such blockage events.
Secondly, HetSLAgg learns the model for predicting received power values in a data driven manner, and thus, it is easily expected that HetSLAgg can learn a feasible model for the prediction even under multiple pedestrians without modifying the aforementioned system architecture as long as a data set containing informative features for the prediction is available.
Hence, the problem boils down to how to obtain such data set, which is beyond the scope of this experiment.

It should be also noted that motivated by our focus on the variation of received powers due to moving obstacles, we arranged the measurement such that the TX and RX do not perform beam tracking.
The TX and DD are equipped with beam tracking, and when the received power at DD is varied, the beam directions of the TX and DD are altered.
Meanwhile, in the measurement, the pedestrian travels between the DD and RX indicated in Fig.~\ref{fig:experimental_environment}, wherein the received power at DD is not altered.
In this situation, the beam direction of the TX and DD is almost fixed. 
Moreover, RX is equipped with a fixed horn antenna and hence does not perform beam tracking.

\begin{figure}[]
	\centering
	\includegraphics[width=\columnwidth]{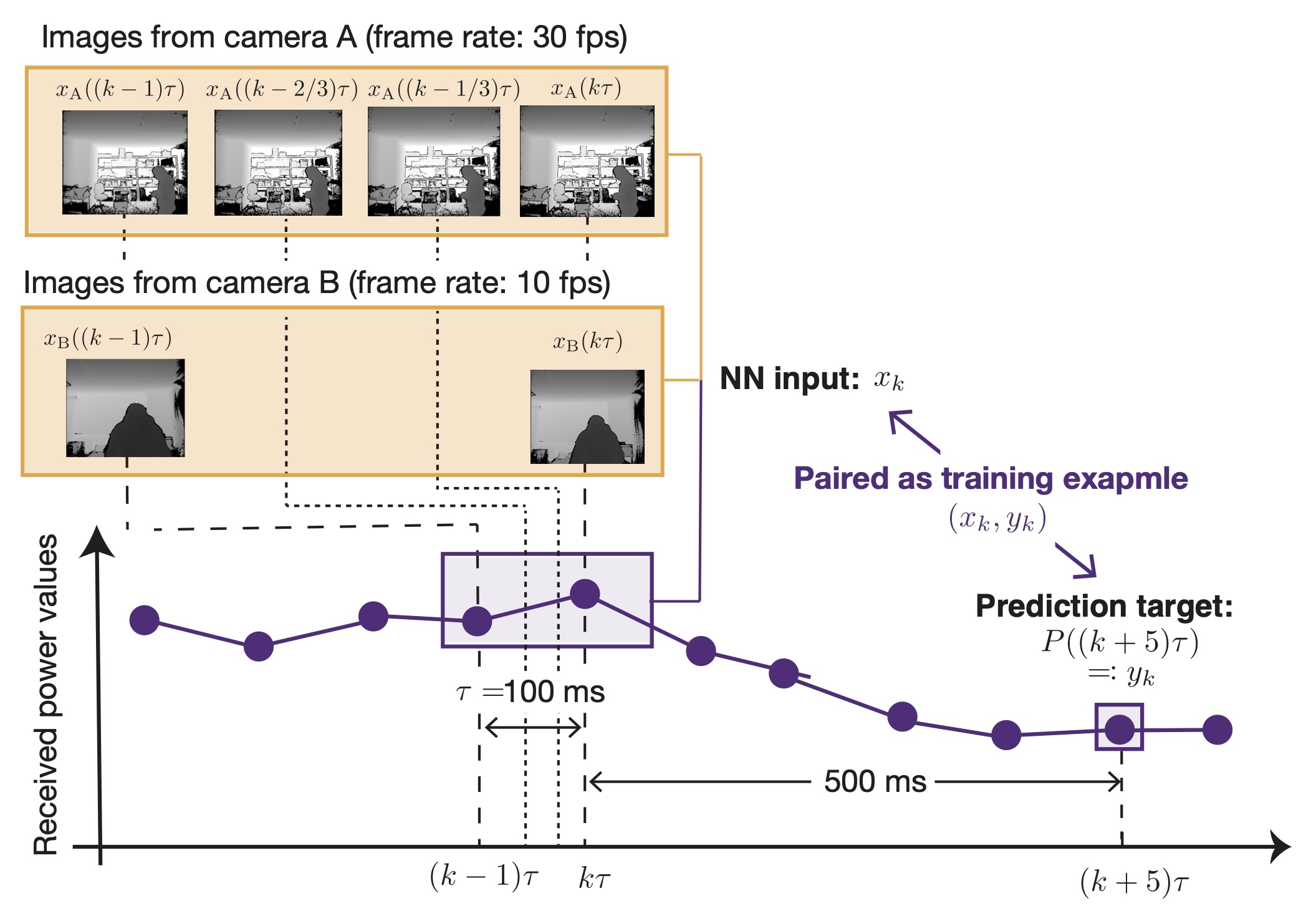}
	\caption{Procedure for creating data set. The images and received powers within past 100\,ms are treated as NN inputs and are denoted as $x_k$.
	The NN inputs are labeled with the received power 500\,ms ahead $P((k + 5)\tau)$, which is treated as a prediction target and denoted as $y_k$.}
	\label{fig:data_set_creation}
\end{figure}

Based on the measured received powers and depth images, we create the data set following the procedure depicted in Fig.~\ref{fig:data_set_creation}.
As discussed above, the frame rate of camera~B is 10\,fps, and hence, the obtained image sequence in camera~B is $\bm{x}^{\mathrm{(B)}}(0), \bm{x}^{\mathrm{(B)}}(\tau), \bm{x}^{\mathrm{(B)}}(2\tau), \dots, \bm{x}^{\mathrm{(B)}}((K - 1)\tau)$, where $\tau = 100\,\mathrm{ms}$, and $K = 3840$ is the total number of image frames obtained through the measurement.
Meanwhile, the frame rate of camera~A is three times higher than that of camera~B, and hence, the obtained image sequence in camera~A within the time duration $[0, (K - 1)\tau]$ is $\bm{x}^{\mathrm{(A)}}(0), \bm{x}^{\mathrm{(A)}}(\tau/3), \bm{x}^{\mathrm{(A)}}(2\tau/3), \dots, \bm{x}^{\mathrm{(A)}}((K - 1)\tau)$.

The prediction is made with past image and received power sequences within the look-back time of $T_{\mathrm{back}} = 100\,\mathrm{ms} =  \tau$ to predict the received power with the look-ahead time of $T = 500\,\mathrm{ms} = 5\tau$, and hence, as the data set, past image and received power sequences within the time-duration of 100\,ms obtained from both camera~A and B are labeled with the received power 500\,ms ahead as shown in Fig.~\ref{fig:data_set_creation}.
Specifically, for $k\in\{1, 2, \dots, K\}$, the images within $[(k - 1)\tau, k\tau]$, i.e., $x_k \coloneqq (\bm{x}^{\mathrm{(B)}}((k - 1)\tau), \bm{x}^{\mathrm{(B)}}(k\tau),  \bm{x}^{\mathrm{(A)}}((k - 1)\tau) ,\bm{x}^{\mathrm{(A)}}((k - 2/3)\tau), \bm{x}^{\mathrm{(A)}}((k - 1/3)\tau), \bm{x}^{\mathrm{(A)}}(k\tau), P((k - 1)\tau), P(k\tau))$ are labeled with the received power $y_k\coloneqq P((k + 5)\tau)$ as depicted in Fig.~\ref{fig:data_set_creation}.
To summarize, the created data set is $\{\,(x_k, y_k)\mid\,k \in\{1, 2, \dots, K\}\}$, over which the training and performance test is performed.

\vspace{3pt}\noindent \textbf{Training and Test.}\quad
The training and test are performed with datasets that differ from each other, which is a common procedure in building ML models\cite{goodfellow}.
Let $k\in \{1, 2, \dots, K\}$ denote the time-index.
We perform training and test with samples whose time-index is in the index set $\mathcal{K}_{\mathrm{train}}$ and $\mathcal{K}_{\mathrm{test}}$, respectively, wherein $\mathcal{K}_{\mathrm{train}}\cup \mathcal{K}_{\mathrm{test}} = \{1, 2, \dots, K\}$, and $\mathcal{K}_{\mathrm{train}}\cap \mathcal{K}_{\mathrm{test}} = \emptyset$.
In this evaluation, the ratio of $|\mathcal{K}_{\mathrm{train}}|$ and $|\mathcal{K}_{\mathrm{test}}|$ is set as 75\% and 25\%, respectively; hence, $\mathcal{K}_{\mathrm{train}} = \{1, 2, \dots, 3K/4\}$ and $\mathcal{K}_{\mathrm{test}}= \{3K/4 + 1, \dots, K\}$.

The training is performed so that the mean square error (MSE) between the predicted and actual received powers is minimal.
Let $f(\cdot;\bm{\theta})$ denotes the NN model with the weight parameters $\bm{\theta}$, and let $\hat{y}_{k}^{\bm{\theta}}\coloneqq f(x_k;\bm{\theta})$ is the prediction of $y_k$.
The parameter is learned by solving the following optimization problem:\vspace{-1em}
\begin{mini}
	{\substack{\bm{\theta}}}
	{\frac{1}{|\mathcal{K}_{\mathrm{train}}|}\sum_{k\in \mathcal{K}_{\mathrm{train}}}\bigl(\hat{y}_k^{\bm{\theta}} -y_k\bigr)^2}{}{},
\end{mini}
The problem is solved with the Adam optimizer\cite{sutskever2013importance} with the learning rate of $1.0\times 10^{-3}$, the decaying rate parameters $\beta_1=0.9$ and $\beta_2=0.999$, and the batch size of 64.
The training is continued until 40 training epochs (1800 stochastic gradient descent steps) are iterated.
Both cameras and BS train their NN layers, i.e., perform forward and backward calculations in parallel computing via exploiting an Nvidia Tesla P100-PCIE GPU with 2560 cores with memory corresponding to 16\,GB and memory bandwidth corresponding to 320\,GB/s.

\begin{table*}
	\centering
	\caption{Split NN Architectures}
	\label{table:nn_arc}
	\begin{tabular}{ccccc}\toprule
		\multicolumn{5}{c}{(a) NN Layers in camera~$i$}\\
		Layer & Filter size & Input shape & Output shape & Remarks\\\midrule
		\texttt{conv1}& 3x3 & 1x40x40x$N^{(i)}_{\mathrm{img}}$ & 64x40x40x$N^{(i)}_{\mathrm{img}}$ & Zero padding, 64 filters\vspace{.2em}\\
		\texttt{norm1} (Batch normalization) & - & 64x40x40x$N^{(i)}_{\mathrm{img}}$ & 64x40x40x$N^{(i)}_{\mathrm{img}}$ & -\vspace{.2em}\\
		\texttt{conv2}& 3x3 & 64x40x40x$N^{(i)}_{\mathrm{img}}$ & 64x40x40x$N^{(i)}_{\mathrm{img}}$ & Zero padding\, 64 filters \vspace{.2em}\\
		\texttt{norm2} (Batch normalization) & - & 64x40x40x$N^{(i)}_{\mathrm{img}}$ & 64x40x40x$N^{(i)}_{\mathrm{img}}$ & -\vspace{.2em}\\
		\texttt{pool} (Avarage pooling) & - & 64x40x40x$N^{(i)}_{\mathrm{img}}$ & {64x20x20x$N^{(i)}_{\mathrm{img}}$} & pooling dimension 2x2, stride: 1\vspace{.2em}\\
		\texttt{recurrent1} (Conv. LSTM)& {3x3} & {64x20x20x$N^{(i)}_{\mathrm{img}}$} & {20x20x$N^{(i)}_{\mathrm{img}}$} & Zero padding\\\bottomrule
	\end{tabular}\\\vspace{1em}
	\begin{tabular}{ccccc}\toprule
		\multicolumn{5}{c}{(b) NN Layers in BS}\\
		Layer & Filter size & Input shape & Output shape & Remarks\\\midrule
		\texttt{recurrent2} (Conv. LSTM) & {3x3} & {20x20x$N_{\mathrm{rss}}$} & {20x20x$N_{\mathrm{rss}}$} & Zero padding\vspace{.2em}\\
		\texttt{Manifold Mixup} & - & {20x20x$N^{(\mathrm{B})}_{\mathrm{img}}$} & {20x20x$N^{(\mathrm{A})}_{\mathrm{img}}$} & Only in \textsf{MmixInt}\vspace{.2em}\\
		\texttt{Aggregation} & - & {2x20x20x$N^{(\mathrm{A})}_{\mathrm{img}}$} & 20x20x$N^{(\mathrm{BS, agg})}_{\mathrm{img}}$ & -\vspace{.2em}\\
		{\texttt{fc1}} & - & 20x20x$N^{(\mathrm{BS,agg})}_{\mathrm{img}}$+20x20x$N_{\mathrm{rss}}$ & 96 & -\vspace{.2em}\\
		\texttt{fc2} & - & 96 & 1 & -\\\bottomrule
	\end{tabular}
\end{table*}


\vspace{3pt}\noindent \textbf{NN Architecture.}\quad
The split NN architecture under study is summarized in Table~\ref{table:nn_arc}.
In Table~\ref{table:nn_arc}, $N_{\mathrm{img}}^{(i)}$ and $N_{\mathrm{rss}} = 2$ denote the lengths of image inputs at camera~$i$ and received power values, respectively, and $N_{\mathrm{img}}^{(\mathrm{BS,agg})}$ denotes the length of aggregated image feature activations.
The length $N_{\mathrm{img}}^{(i)}$ depends on methods for balancing frame rate difference (i.e., \textsf{Disc}, \textsf{MixInt}, and \textsf{MmixInt}).
Specifically, $N_{\mathrm{img}}^{\mathrm{(A)}} = N_{\mathrm{img}}^{\mathrm{(B)}} = 2$ in \textsf{Disc}, $N_{\mathrm{img}}^{\mathrm{(A)}} = N_{\mathrm{img}}^{\mathrm{(B)}} = 4$ in \textsf{MixInt}, and $N_{\mathrm{img}}^{\mathrm{(A)}} = 4$ and $N_{\mathrm{img}}^{\mathrm{(B)}} = 2$ in \textsf{MmixInt}.
The length $N_{\mathrm{img}}^{(\mathrm{BS,agg})}$ depends on the aggregation methods for image feature activations (i.e., \textsf{ConcAgg} and \textsf{MmixAgg}).
Specifically, $N_{\mathrm{img}}^{(\mathrm{BS,agg})} = 2N_{\mathrm{img}}^{\mathrm{(A)}}$ in \textsf{ConcAgg} whereas  $N_{\mathrm{img}}^{(\mathrm{BS,agg})} = N_{\mathrm{img}}^{\mathrm{(A)}}$ in \textsf{MmixAgg}.
Note that the choice of the layer stack and parameters is consistent with the prior works \cite{nishio_jsac} and \cite{koda2020communication}, which addresses a similar training task.
Specifically, the choice of the layer stack and parameters at the cameras in Table~\ref{table:nn_arc}(a) are consistent with \cite{nishio_jsac}, which achieves feasible prediction accuracy in mmWave received power prediction based on visual data.
Moreover, except for the layers particular in this work, e.g., \texttt{Manifold Mixup} and \texttt{Aggregation} layers, the choice of the layer stack and parameters at the BS in Table~\ref{table:nn_arc}(b) is consistent with \cite{koda2020communication}, which was shown to successfully fuse RF and visual data.

Each camera involves two two-dimensional convolutional layers termed \texttt{conv1} with 64 filters with the size of 3x3 and \texttt{conv2} with 64 filters with the size of 3x3.
Being consistent with \cite{nishio_jsac}, each camera feeds the outputs of \texttt{conv1} and \texttt{conv2} into batch normalization layers termed \texttt{norm1} and \texttt{norm2}, respectively, which allows us to less careful about the initialization of NN parameters to accelerate training\cite{ioffe2015batch}.
The outputs from \texttt{norm2} is fed into the average pooling procedure termed \texttt{pool}, which reduces the feature dimension and thereby reducing both the communication cost to upload the features to BS and the energy cost to perform calculations in the subsequent layers.
The outputs of \texttt{pool} are fed into the recurrent layer termed \texttt{recurrent1}.
The \texttt{recurrent1} is implemented with convolutional long-term short memory\cite{xingjian2015convolutional} that allows us to obtain spatial and temporal features of input images, which was shown to successfully achieve better performance than other recurrent procedures in a same training task\cite{nishio_jsac}.

The BS is implemented with one recurrent layer \texttt{recurrent2}, one aggregation layer \texttt{aggregation}, and two fully connected layers termed \texttt{fc1} and \texttt{fc2}.
Specifically in \textsf{MmixInt}, the BS is also implemented with one Manifold Mixup layer termed \texttt{Manifold Mixup}.
First, the BS feeds received power values into \texttt{recurrent2} and calculates RF feature activations.
In a parallel way, the \texttt{Aggregation} performs the aggregation of uploaded image feature activations with either method discussed in Section~\ref{subsec:aggregation}.
Subsequently, the RF and aggregated image feature activations are fed into \texttt{fc1} with 96 hidden units.
Finally, feeding the hidden activations of \texttt{fc1} into \texttt{fc2}, the BS calculates the predicted value of a future received power.

\subsection{Communication Channel Model}
Communications between each camera and BS are performed over wireless channels.
Specifically, each camera and BS communicate to exchange forward/backward propagation signals in both proposed HetSLAgg and HetSLFedAvg baseline as shown in Fig.~\ref{fig:system_model}.
In what follows, we specify the wireless channel and transmission rate.

\vspace{3pt}\noindent \textbf{Channel Between Camera A and BS.}\quad
Camera~A and BS communicate over the mmWave channel specified in Fig.~\ref{fig:experimental_environment}.
As discussed above, a pedestrian causes a blockage event periodically, and the received power varies according to the pedestrian movement.
Therein, the received power in a blockage event is smaller than that in a LoS condition by approximately 15\,dB.

Accordingly, the transmission rate between camera A and BS varies over time.
Let the subscript $(i, j)\in\{(\mathrm{A}, \mathrm{BS}), (\mathrm{BS}, \mathrm{A})\}$ indicate the pair of the destination and source node. 
The transmission rate from the source node~$i$ to the destination node~$j$ is given by:
\begin{align}
	\label{eq:trans_rate_A_BS}
	R_{i, j}(t) = W \log_2\!\left(1 + \frac{A(t) P_{\mathrm{L},i, j}}{N}\right)
\end{align}
where $P_{\mathrm{L},i,j}$ is the power of the signal transmitted from node~$i$ received at node~$j$ in a LoS condition and measured as $P_{\mathrm{L}, i, j} = -29$\,dBm $\forall (i, j)$ in the aforementioned experiment.
The values $W = 1.76$\,GHz and $N = -60$\,dBm denote the bandwidth and noise power, respectively.
In \eqref{eq:trans_rate_A_BS}, $A(t)$ is the time-varying received power attenuation value.
We measured the attenuation values periodically (per $\tau = 100$\,ms) in the aforementioned experiment, and hence, we represent $A(t)$ as a staircase function using measured attenuation values: $A(t) = \sum_{k = 0}^{K}A_{\mathrm{mes}, k}\mathbbm{1}_{\{t\in[k\tau, (k + 1)\tau)\}}$, where $(A_{\mathrm{mes}, k})_{k\in\{0, 1, \dots, K\}}$ denote the measured attenuation value at $t = 0, \tau, \dots, K\tau$, and $\mathbbm{1}_{\{\mathit{condition}\}}$ denotes the indicator function that equals 1 if $\mathit{condition}$ is satisfied and 0, otherwise.
Note that from the above caluclation, both $R_{\mathrm{A}, \mathrm{BS}}$ and $R_{\mathrm{BS}, \mathrm{A}}$ ranges from approximately 5\,Gbit/s to 18\,Gbit/s, which is within the data rate supported by the IEEE 802.11ay standard\cite{ghasempour2017ieee} and is not an unrealistic value.

\vspace{3pt}\noindent \textbf{Channel Between Camera B and BS.}\quad
Similarly, camera~B communicates with the BS over the same channel band as camera~A to BS.
Unlike the link between camera~A and BS, the channels are not blocked by the pedestrian as illustrated in Fig~\ref{fig:experimental_environment}.
Hence, the transmission rate between camera~B and BS does not vary over time.
Let $(i', j')\in\{(\mathrm{B}, \mathrm{BS}), (\mathrm{BS}, \mathrm{B})\}$ indicate the pair of destination node and  source node. 
The transmission rate from the source node~$i'$ to the destination node~$j'$ is given by:
\begin{align}
	\label{eq:trans_rate_B}
	R_{i', j'} = W \log_2\!\left(1 + \frac{P_{\mathrm{L}, i', j'}}{N}\right).
\end{align}
We calculate $P_{\mathrm{L}, i', j'}$ from the power-distance law, which is well-accepted and was shown to match actual received powers particularly in LoS conditions\cite{geng2008millimeter}.
Following the law, we calculate the power in a LoS condition as:
\begin{align}
	10\log_{10}\!\left(P_{\mathrm{L}, i', j'}\right) 
	&= 10\log_{10}\!\left(P_{\mathrm{T}, i'}\right) + 10\log_{10}\!\left(G_{i'}G_{j'}\right)\nonumber\\
	&\quad - \mathrm{PL}_0(d_0) - 10n\log_{10}\!\left(\frac{d_{\mathrm{B}, \mathrm{BS}}}{d_0}\right), 
\end{align}
where $10\log_{10}(P_{\mathrm{T}, i'})$ is the transmit power in dB scale and set as $10$\,dBm, $d_{\mathrm{B}, \mathrm{BS}}$ is the distance between camera~B and BS, and $n$ is the path loss exponent and is set as 1.6\cite{geng2008millimeter}.
The gain $\mathrm{PL}_0(d_0)$ is the path loss at reference distance $d_0$ in dB scale and is 68\,dB with  $d_0 = 1$\,m at 60\,GHz\cite{geng2008millimeter}.
The value $10\log_{10}(G_{i'})$ is the antenna gain of the node~$i'$, and we set $10\log_{10}(G_{\mathrm{BS}}) = 8$\,dBi, and $10\log_{10}(G_{\mathrm{B}}) = 24$\,dBi.
Note that from the above caluclation, both $R_{\mathrm{B}, \mathrm{BS}}$ and $R_{\mathrm{BS}, \mathrm{B}}$ result in approximately 19\,Gbit/s, which is again within the data rate supported by the IEEE 802.11ay standard.

\subsection{Performance Metrics}

\vspace{3pt}\noindent \textbf{Prediction Accuracy.}\quad
Prediction accuracy is evaluated using the RMSE.
Given the predicted received powers $(\hat{y}^{\bm{\theta}}_k)_{k\in\mathcal{K}_{\mathrm{test}}}$ in the trained parameters ${\bm{\theta}}$, the RMSE is given as follows:
\begin{align}
	\text{RMSE} = \sqrt{\frac{\sum_{k\in\mathcal{K}_{\mathrm{test}}}(\hat{y}^{\bm{\theta}}_k - y_k)^2}{|\mathcal{K}_{\mathrm{test}}|}}.
\end{align}


\vspace{3pt}\noindent \textbf{Forward and Backward Propagation Latency.}\quad
The latency for transmitting forward and backward propagation signals is calculated as follows.
Cameras~A and~B operates at the same channel band, and hence, they communicate with the BS in a time-division fashion to avoid co-channel interference.
Let the shorthand notations $[k]$ denote the interval $[(k - 1)\tau, k\tau]$ for $k\in\{1, 2, \dots, K\}$.
The latency for transmitting forward propagation signals within the interval $[k]$ is denoted by $T_{\mathrm{FP}}[k]$ and is calculated as follows:
\begin{align}
	\label{eq:forward_time}
	T_{\mathrm{FP}}[k] = U(\mathrm{A})\frac{D_{\mathrm{FP}}^{(\mathrm{A})}}{R_{\mathrm{A, BS}}(k\tau)} + U(\mathrm{B})\frac{D_{\mathrm{FP}}^{(\mathrm{B})}}{R_{\mathrm{B, BS}}},
\end{align}
where $D_{\mathrm{FP}}^{(i)}$ for $i\in\{\mathrm{A}, \mathrm{B}\}$ is the data size of forward propagation and corresponds to the image feature activation uploaded from camera~$i$.
Denoting $\bm{a}^{(i)}$ as the feature activation uploaded from camera~$i$ consistently with Section~\ref{sec:manifold_mixup}, the data side $D_{\mathrm{FP}}^{(i)}$ is calculated by multiplying $\dim(\bm{a}^{(i)})$ by the bit resolution $r = 32$, i.e., $D_{\mathrm{FP}}^{(i)} = r\dim(\bm{a}^{(i)})$.
The payload size $D_{\mathrm{FP}}^{(i)}$ depends on methods for balancing frame rate difference, i.e., \textsf{Disc}, \textsf{MixInt}, and \textsf{MmixInt}.
Specifically, $D_{\mathrm{FP}}^{(\mathrm{A})} = D_{\mathrm{FP}}^{(\mathrm{B})} = 3.2$\,kB in \textsf{Disc}, $D_{\mathrm{FP}}^{(\mathrm{A})} = D_{\mathrm{FP}}^{(\mathrm{B})} = 6.4$\,kB in \textsf{MixInt}, and $D_{\mathrm{FP}}^{(\mathrm{A})} = 3.2$\,kB and $D_{\mathrm{FP}}^{(\mathrm{B})} = 6.4$\,kB in \textsf{MmixInt}.
In \eqref{eq:forward_time}, $U(i)$ denote whether BS exchanges the forward and backward propagation signals in camera~$i$ or not and equals 1 if BS exchanges the signals with camera~$i$ and 0, otherwise.
For example, in the proposed HetSLAgg, $U(\mathrm{A})$ and $U(\mathrm{B})$ always equal 1 because BS exchanges the forward and backward propagation signals with both camera~A and camera~B as shown in Fig.~\ref{fig:system_model}(a).
On the contrary, in the HetSLFedAvg baseline, either $U(\mathrm{A})$ or $U(\mathrm{B})$ equals 1 because BS exchanges the forward and backward propagation signals with either camera~A or camera~B as shown in Fig.~\ref{fig:system_model}(b).
Likewise, the latency for transmitting back propagation signals within the interval $[k]$ is denoted by $T_{\mathrm{BP}}[k]$ and is calculated as follows:
\begin{align}
	\label{eq:backward_time}
	T_{\mathrm{BP}}[k] = U(\mathrm{A})\frac{D_{\mathrm{BP}}}{R_{\mathrm{BS, A}}(k\tau)} +  U(\mathrm{B})\frac{D_{\mathrm{BP}}}{R_{\mathrm{BS, B}}},
\end{align}
where $D_{\mathrm{BP}}$ denote the data size of backward propagation and corresponds to the dimension of the gradient in $\texttt{fc1}$ layer, i.e., $D_{\mathrm{BP}} = r\dim(\nabla \bm{\theta}_{\texttt{fc1}})$, where $\bm{\theta}_{\texttt{fc1}}$ is the weight parameters in \texttt{fc1} layer.
The payload size $D_{\mathrm{BP}}$ depends on both interpolation and image feature aggregation methods.
Specifically, in \textsf{MixInt} and \textsf{MmixInt}, $D_{\mathrm{BP}} = 1843.2$\,kB in \textsf{ConcAgg} whereas $D_{\mathrm{BP}} = 1228.8$\,kB in \textsf{MmixAgg}.
In \textsf{Disc}, $D_{\mathrm{BP}} = 921.6$\,kB in \textsf{ConcAgg} whereas $D_{\mathrm{BP}} = 614.4$\,kB in \textsf{MmixAgg}.

We also calculate the total time duration during which a forward and backward propagation signal exchange is completed to evaluate the training time as discussed in the next section.
Let the time duration for completing a forward and backward propagation signal exchange within interval $[k]$ be denoted by $T_{\mathrm{tot}}[k]$, which is calculated by:
\begin{align}
	T_{\mathrm{tot}}[k] = T_{\mathrm{FP}}[k] + T_{\mathrm{BP}}[k] + T_{\mathrm{comp}},
\end{align}
where $T_{\mathrm{comp}}$ is the sum of the time length for calculating the forward and backward propagation signals, which is obtained by measuring the time-duration during which the GPU computes the forward and backward propagations.

\vspace{3pt}\noindent \textbf{Tranining Time.}\quad
To evaluate the training speed, we calculate the time elapsed until the $n$th forward and backward propagation signal exchange is performed and plot its corresponding test accuracy.
Let $N[k]\coloneqq \lfloor\tau/T_{\mathrm{tot}}[k]\rfloor$ denote the maximum number of the forward and backward propagation signal exchange performed within $[k]$.
The $n$th forward and backward propagation exchange is performed in a certain interval, whose index is denoted as $k_n$ and is given by $\max\,\{\,k'\mid \sum_{k = 1}^{k'}N[k] \leq n\,\}$.
We calculate the time elapsed until the $n$th forward and backward propagation exchange is performed, denoted by $T_n$, as follows:
\begin{align}
	T_n = \sum_{k = 1}^{k_n - 1}T_{\mathrm{tot}}[k] + \!\left(n - \sum_{k = 1}^{k_n - 1} N [k] + 1\right)T_{\mathrm{tot}}[{k_n}].
\end{align}

\vspace{3pt}\noindent \textbf{Power Consumption for Operating NN Layers.}\quad
The power consumption in operating the NN layers in the BS and cameras are measured by the total energy costs required for the NN layers performing the addition and multiply calculations and memory access to load the weight parameters\cite{han2015learning}.
Let $E_{\mathrm{add}}$, $E_{\mathrm{mult}}$, and $E_{\mathrm{access}}$ denote the energy required for addition calculations, multiply calculations, and memory access to load one weight parameter, respectively.
The power consumption $P$ operating the NN layers is calculated by:
\begin{align}
	P = \frac{N_{\mathrm{add}} E_{\mathrm{add}} + N_{\mathrm{mult}} E_{\mathrm{mult}} + N_{\mathrm{param}} E_{\mathrm{access}}}{\tau},
\end{align}
where $N_{\mathrm{add}}$, $N_{\mathrm{mult}}$, and $N_{\mathrm{param}}$ are the numbers of addition calculations, multiply calculations, and weight parameters, respectively.
In this evaluation, we set $E_{\mathrm{add}} = 0.9\,\mathrm{pJ}$, $E_{\mathrm{mult}} = 3.7\,\mathrm{pJ}$, and $E_{\mathrm{access}} = 640\,\mathrm{pJ}$ assuming $45\,\mathrm{nm}$ CMOS process as an example\cite{horowitz2014energy}.

\subsection{Effectiveness of Aggregating Image Features in HetSLAgg}

\begin{figure}[t]
	\centering
	\includegraphics[width=0.9\columnwidth]{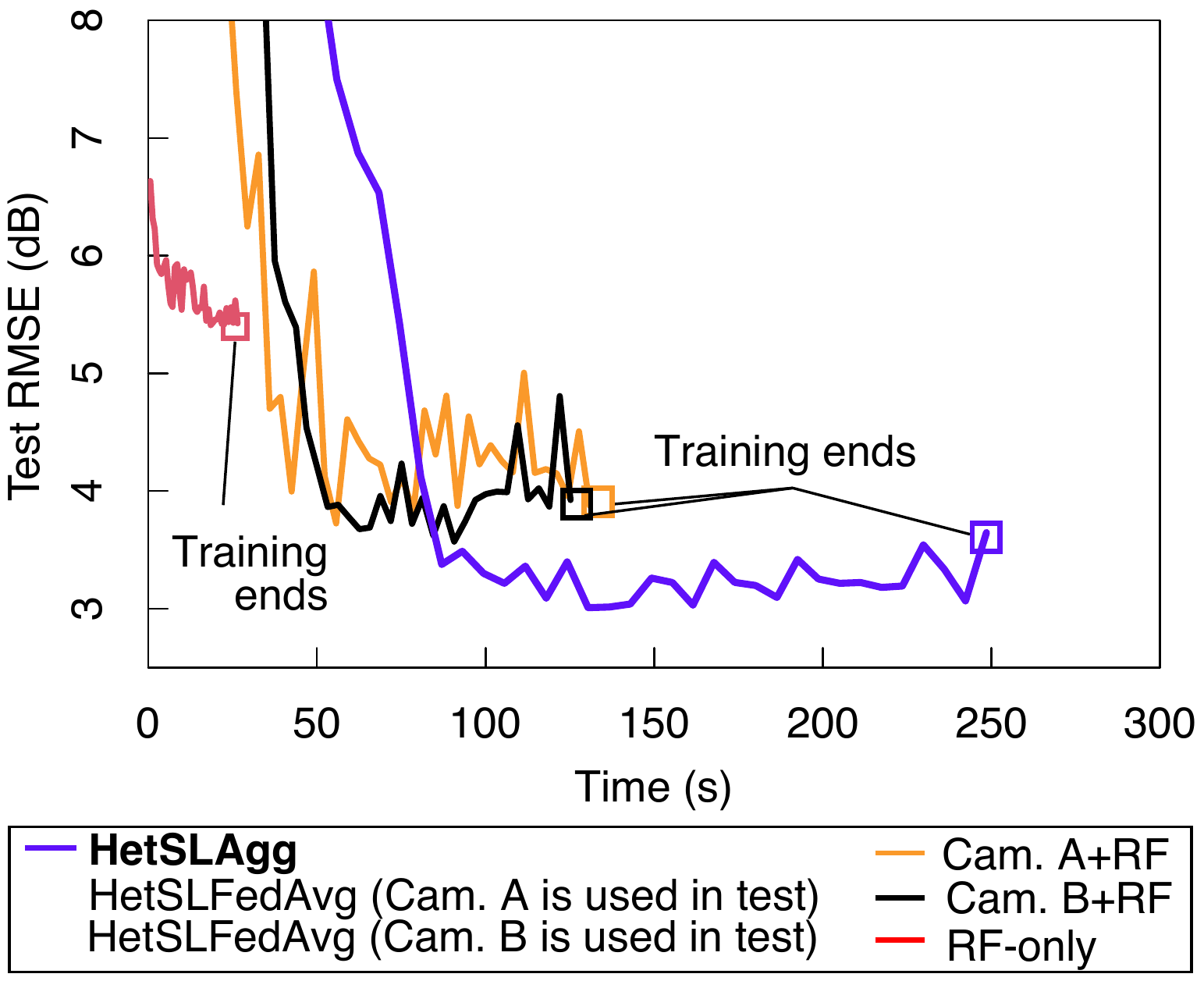}
	\caption{
		Test RMSE w.r.t. time in the proposed HetSLAgg vs. HetSLFedAvg baseline and other baselines, i.e.,  single camera and received power-based predictions and received power-based predictions.
		}
	\label{fig:compare_hetslagg_hetslfedavg}
\end{figure}

\vspace{3pt}\noindent\textbf{Training Time vs. Accuracy.}\quad
First, we validate that the proposed HetSLAgg achieves better prediction accuracy relative to HetSLFedAvg baseline and other baselines, i.e., single camera and received power-based predictions ({Cam.\,A+RF} and {Cam.\,B+RF}) and received power-based predictions ({RF-only}).
In this evaluation, as a balancing method for different frame rates in HetSLAgg, we employed \textsf{Disc} as shown in Fig.~\ref{fig:interpolation}(c) as an example.
In addition, as an image feature aggregation method in HetSLAgg, we employed \textsf{ConcAgg} as shown in Fig.~\ref{fig:aggregation_method}(b) as an example.
In Fig.~\ref{fig:compare_hetslagg_hetslfedavg} showing the test RMSE with respect to the elapsed time in training,  HetSLFedAvg achieves a poorer prediction accuracy relative to not only the proposed HetSLAgg, but also {Cam.\,A+RF} and {Cam.\,B+RF}.
This can be attributed to the fact that the cameras monitor from different angles, resulting in non-IID data distributions between cameras.
On the contrary, the proposed HetSLAgg achieves the best prediction accuracy among the prediction frameworks tested in Fig.~\ref{fig:compare_hetslagg_hetslfedavg}, benefitting from  two diverse image features uploaded from two cameras by fusing their feature activations.

It should be noted that in terms of training speed, the RF-only baseline exhibits faster training followed by Cam.\,A+RF and Cam.\,B+RF baselines completes the training second fastest.
The reason behind the fastest training of RF-only is that the BS does not exchange forward and backward propagation signals with cameras.
In contrast in Cam.\,A+RF and Cam.\,B+RF, BS  exchanges forward and backward propagation signals with one camera, resulting in a lower latency than HetSLAgg and HetSLFedAvg that exchanges forward and backward propagation signals with two cameras.
However, RF-only baseline converges to the test RMSE of approximately 5.4\,dB whereas Cam.\,A+RF and Cam.\,B+RF baselines converge to the test RMSE of approximately 3.6\,dB.
These test RMSEs are higher relative to HetSLAgg that converges to approximately 3.0\,dB, showing the best performance among all baselines.

\begin{figure}[t]
	\centering
	\subfigure[Time series of the received powers predicted 500\,ms prior to the observation of ground truth.]{\includegraphics[width=0.9\columnwidth]{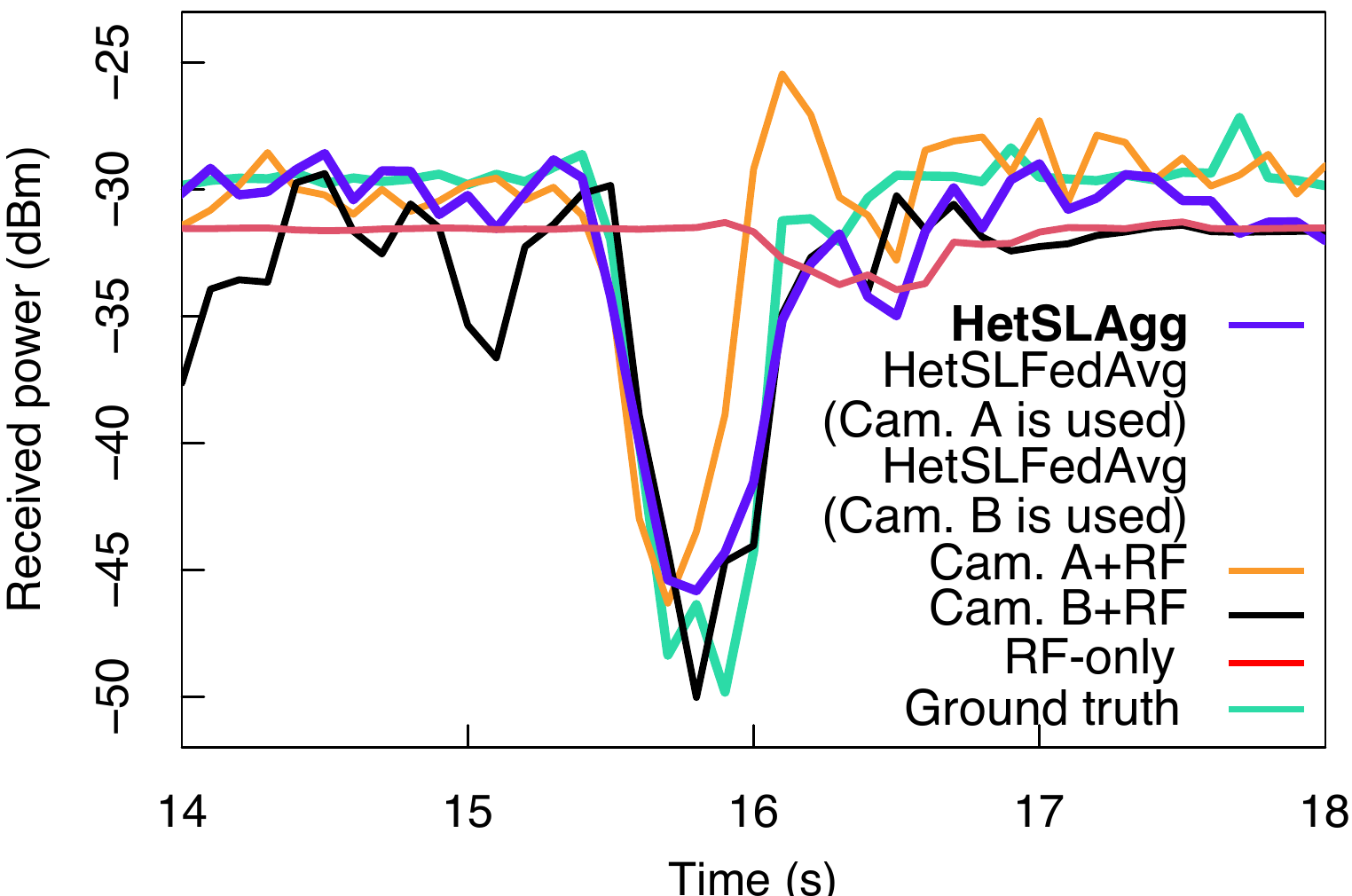}}
	\subfigure[RMSE in different channel conditions.]{\includegraphics[width=0.9\columnwidth]{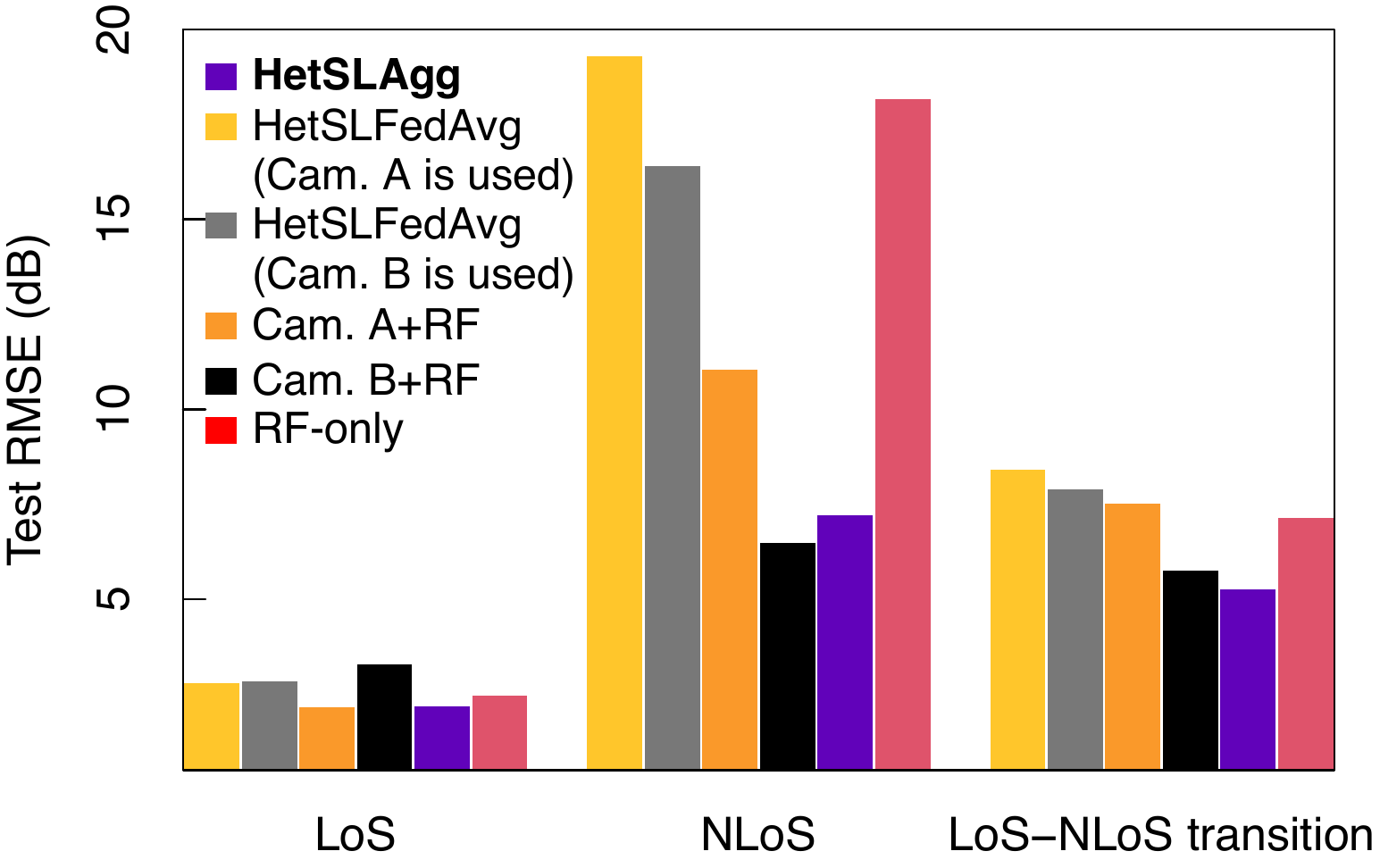}}
	\caption{
		Received power prediction results in test.
		}
	\label{fig:prediction_results}
\end{figure}

\vspace{3pt}\noindent\textbf{Test Accuracy.}\quad
In Fig.~\ref{fig:prediction_results}, we evaluate the prediction performance of the proposed HetSLAgg in detail and show that it outperforms the baselines.
Fig.~\ref{fig:prediction_results}(a) shows the example of the time series of the actual received powers and that of the received powers predicted 500\,ms before the actual powers were observed.
In Fig.~\ref{fig:prediction_results}(a), the baseline of RF-only does not match the ground truth as accurately as the baselines of Cam.\,A+RF and Cam.\,B+RF and the proposed HetSLAgg particularly in the NLoS condition and LoS-NLoS transition conditions. 
In addition, the HetSLFedAvg baseline also does match the ground truth as accurately as Cam.\,A+RF, Cam.\,B+RF, and HetSLAgg owing to the aforementioned problem of the non-IID data distributions among the cameras.
Meanwhile, the predicted received powers in the proposed HetSLAgg match the ground truth better than the examined baselines.
The reason behind the performance of the proposed HetSLAgg is that the HetSLAgg benefits from the two diverse features uploaded from the two cameras.
As observed in the prediction results in Fig.~\ref{fig:prediction_results}(a), Cam\,A+RF performs better in the LoS conditions while Cam\,B+RF performs better in the NLoS and LoS-NLoS conditions.
These facts are quantitatively validated in Fig.~\ref{fig:prediction_results}(b) showing the channel condition-wise RMSE.
Fig.~\ref{fig:prediction_results}(b) demonstrates that the test RMSE in LoS conditions in Cam\,A+RF is lower than the other baselines, and the test RMSE in NLoS and LoS-NLoS transition conditions in Cam\,B+RF is lower than the other baselines.
From Fig.~\ref{fig:prediction_results}(b), we can see that the proposed HetSLAgg takes advantage of Cam\,A+RF and Cam\,B+RF.
Hence, we can conclude that HetSLAgg performs more accurate predictions than the examined baselines benefitting from the two diverse image features obtained from the multiple cameras.

\subsection{Effectiveness of Interpolating and Aggregating Image Features}

\begin{figure}[t]
	\centering
	\includegraphics[width=\columnwidth]{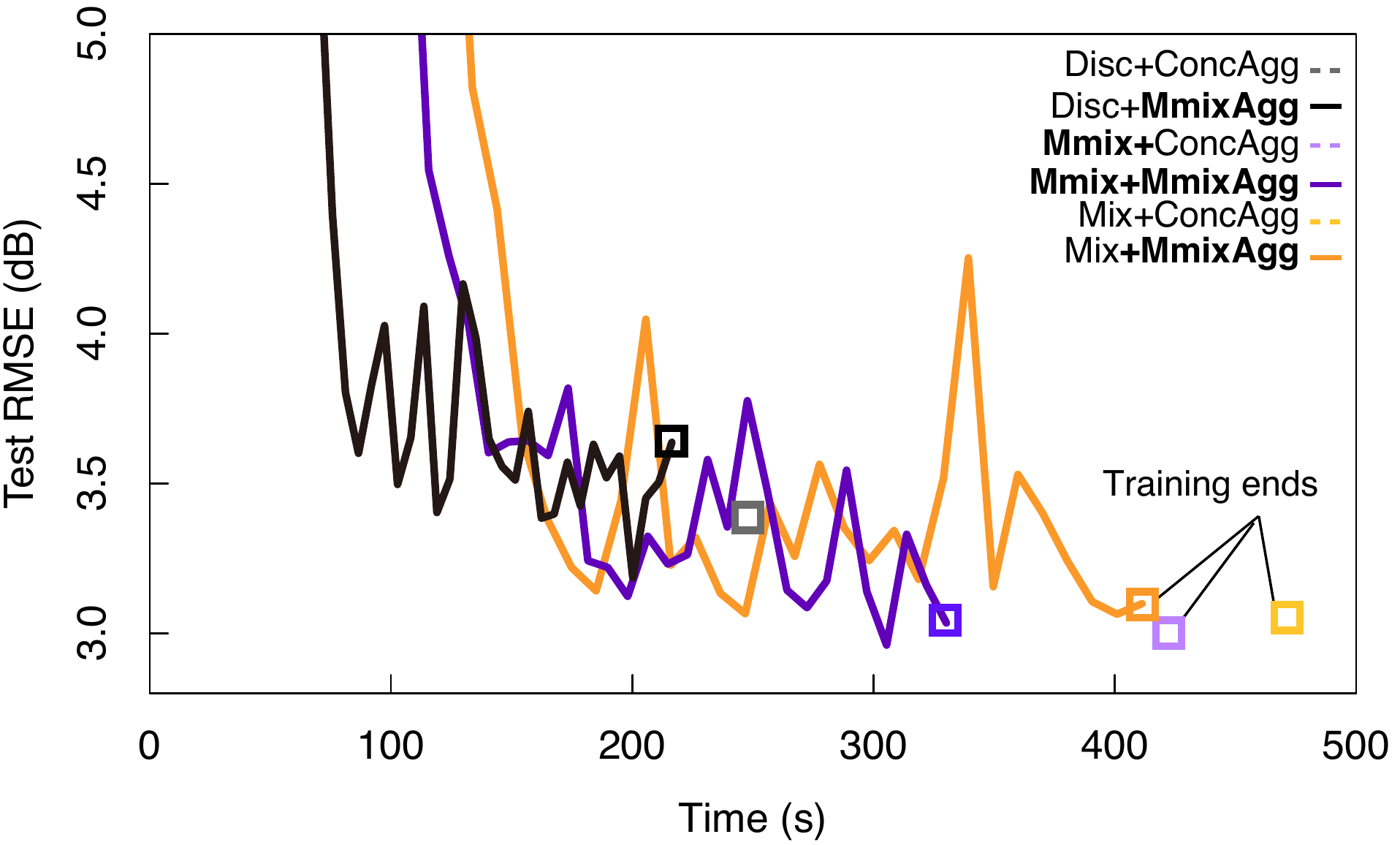}
	\caption{Test RMSE w.r.t. time in the proposed HetSLAgg with different methods for balancing frame rate difference and for aggregating feature activations from two cameras.}
	\label{fig:training_balancing_methods}
\end{figure}

\vspace{3pt}\noindent\textbf{Training Time vs. Accuracy.}\quad
We first compare the three methods for balancing different frame rates, i.e., the \textsf{Disc} baseline, \textsf{MixInt} baseline, and \textsf{MmixInt} in terms of training performance and show that the \textsf{MmixInt} completes the training faster than \textsf{MixInt} while achieving better or comparable prediction accuracies over the two other methods.
Fig.~\ref{fig:training_balancing_methods} shows the test RMSE with respect to the time elapsed in training for different methods for balancing different frame rates and those for aggregating image feature activations. 
In Fig.~\ref{fig:training_balancing_methods}, we can see that the \textsf{Disc} baseline completes the training faster than the \textsf{MixInt} and  \textsf{MmixInt} by the order of 100\,s.
This is due to the lower payload size of forward and backward propagation signals, wherein the exchange of the forward and backward propagation signals can be completed fastest.
On the contrary,  \textsf{MixInt} and  \textsf{MmixInt} complete the training with the lower RMSE relative to the \textsf{Disc}, benefitting from both the images from camera~A with a higher frame rate and interpolated images in camera~B that facilitate capturing the pedestrian movement in a higher time-resolution.
While achieving the comparable test RMSE, the proposed \textsf{MmixInt} completes the training faster than \textsf{MixInt} owing to the reduced payload size for forward propagation signals and consequent a lower latency for exchanging forward and backward propagation signals.

In Fig.~\ref{fig:training_balancing_methods}, we also compare the two feature aggregation methods, i.e., proposed \textsf{MmixAgg} and the \textsf{ConcAgg} baseline in terms of training performance.
In Fig.~\ref{fig:training_balancing_methods}, we can see that \textsf{MmixAgg} completes the training faster than  \textsf{ConcAgg}.
This can be explained by the lower payload size of backward propagation signals in \textsf{MmixAgg}.
More specifically, the payload size of backward propagation signals is proportional to the number of weight parameters in the \texttt{fc1} layer, which could be reduced by aggregating the features with \textsf{MmixAgg} (see. Fig.~\ref{fig:aggregation_method}).
Consequently, this results in a lower latency for transmitting the backward propagation signals, which led to faster training.
Remarkably, besides the discarding method, the combination of \textsf{MmixAgg} and \textsf{MmixInt} completes model  training faster than the other combinations by approximately 100\,s while achieving comparable test RMSE.

\begin{figure*}[t]
	\centering
	\subfigure[Transmission latency for forward propagation signals.]{\includegraphics[width=0.3\textwidth]{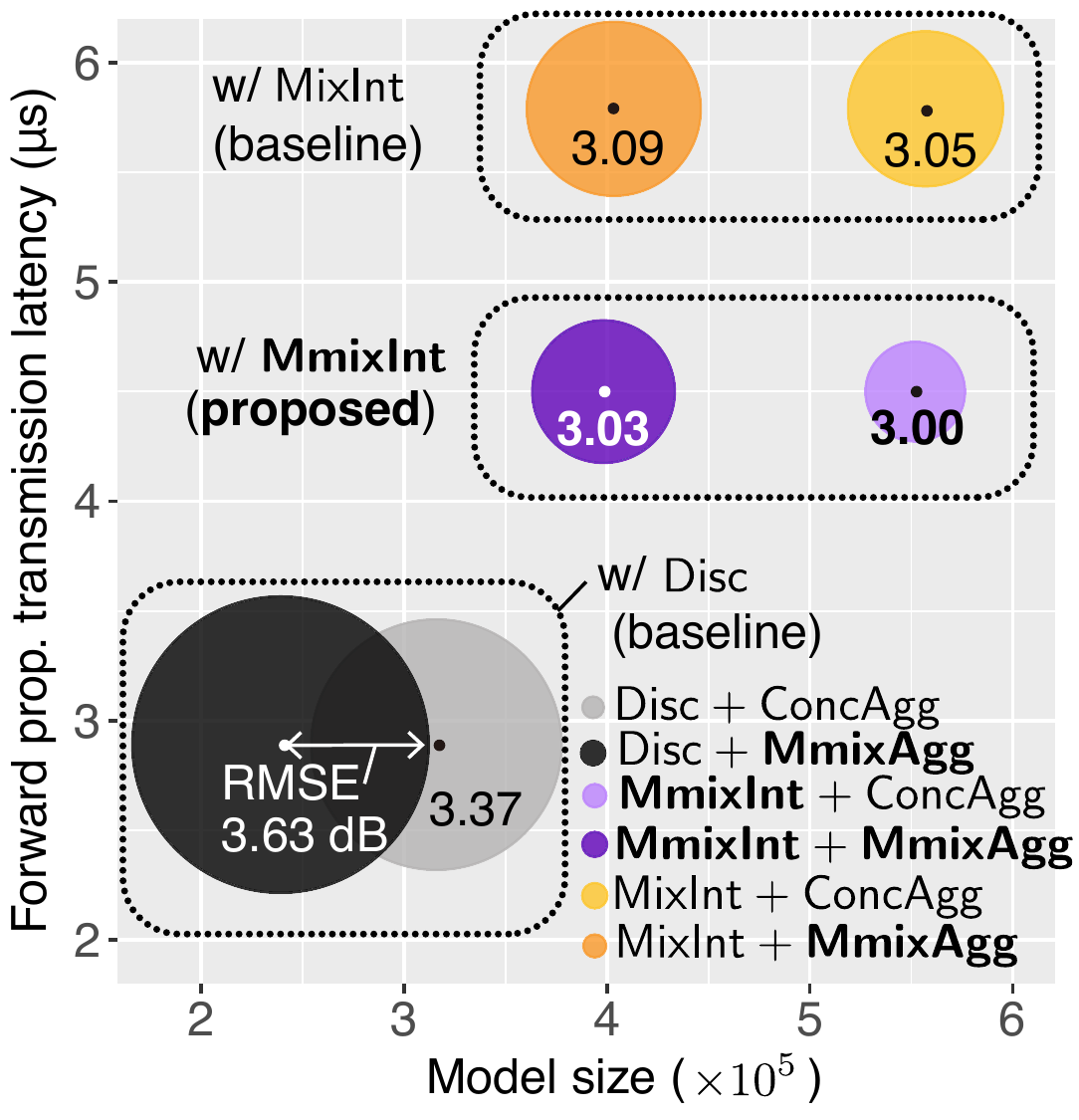}}\hspace{1em}
	\subfigure[Toal power consumption in two cameras.]{\includegraphics[width=0.3\textwidth]{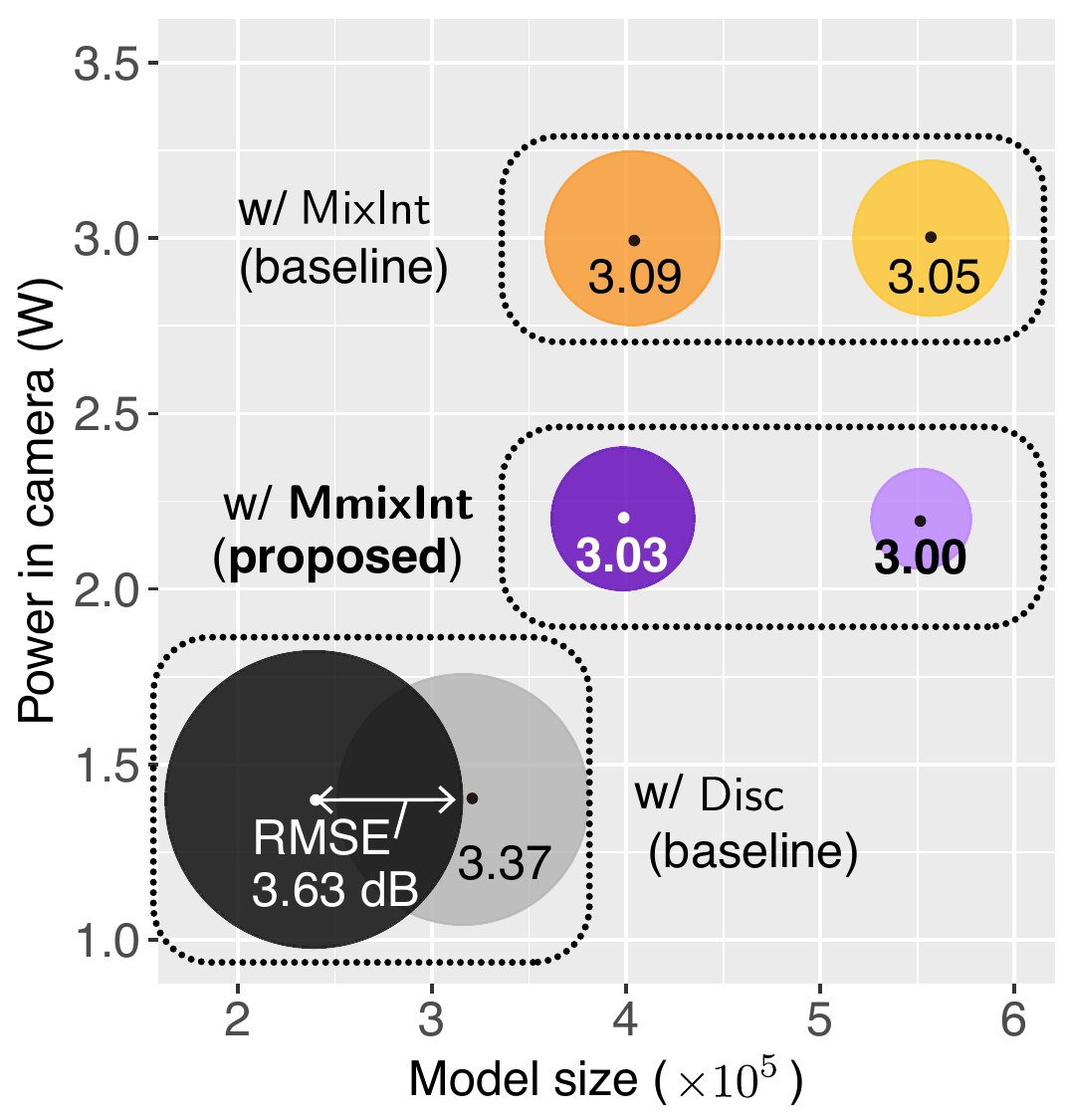}}\hspace{1em}
	\subfigure[Power consumption in BS.]{\includegraphics[width=0.3\textwidth]{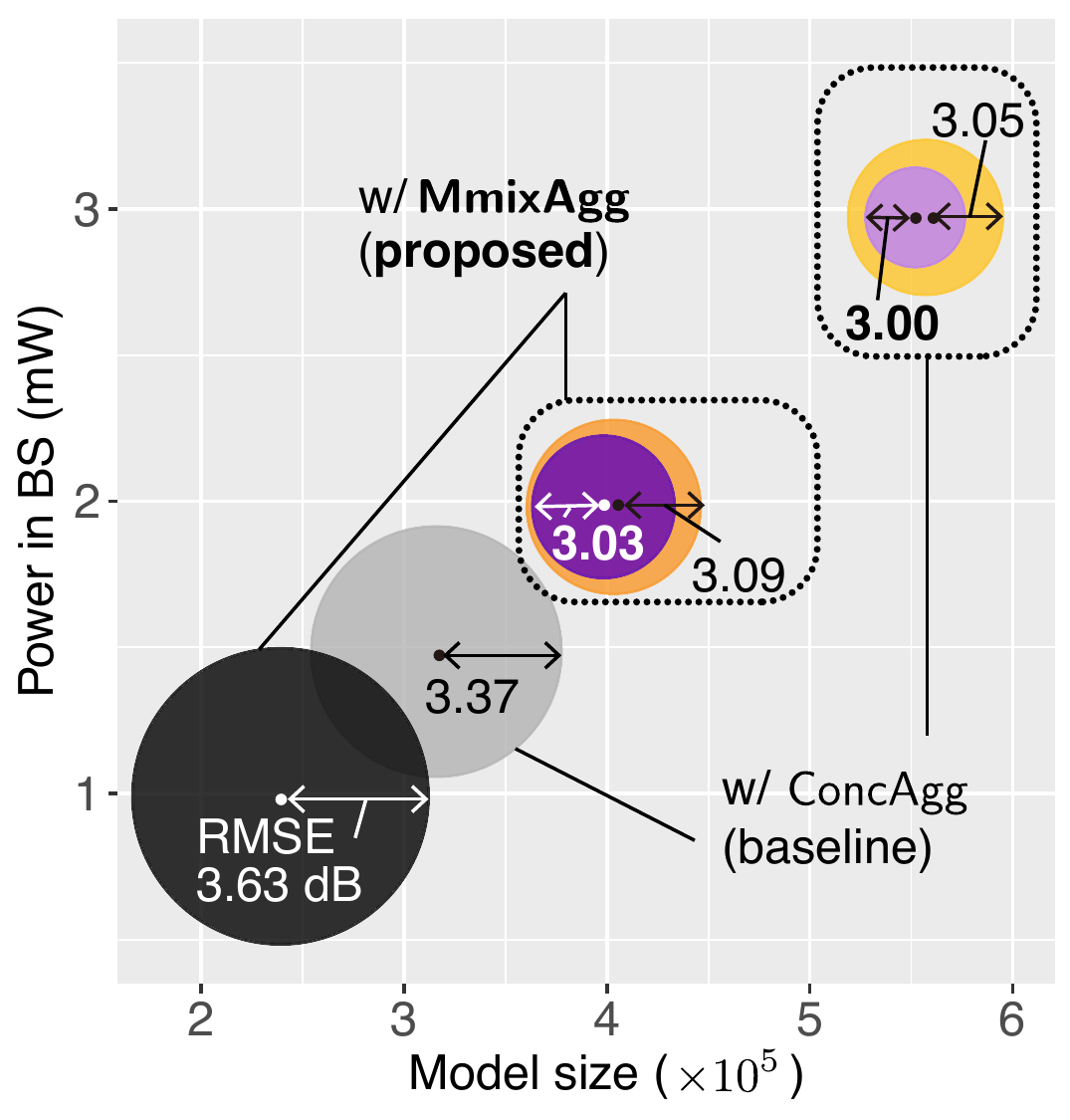}}\hspace{1em}
	\caption{Comparison among balancing methods for different frame rates and feature aggregation methods in different performance metrics.
	The radius of the circles represents the test RMSE, and the smaller circles represent better prediction accuracies.}
	\label{fig:comprehensive_comparison}
\end{figure*}

\vspace{3pt}\noindent\textbf{Test Accuracy and Communication/Energy Efficiency.}\quad
In Figs.~\ref{fig:comprehensive_comparison}(a)--\ref{fig:comprehensive_comparison}(b), we compare the three methods for balancing different frame rates in terms of the three performance metrics, i.e., prediction accuracy, power consumption in calculating the image features in the cameras, and transmission latency for forward propagation signals.
Fig.~\ref{fig:comprehensive_comparison}(a) summarizes the transmission latency for forward propagation signals while Fig.~\ref{fig:comprehensive_comparison}(b) summarizes the total power consumption in the two cameras.
Note that the RMSE is represented by the size of the circles, where the smaller circles represent better prediction accuracies.
First, the \textsf{Disc} baseline is shown to be the most lightweight, i.e., achieves the lowest transmission latency and power consumption owing to the reduced number of images relative to \textsf{MmixInt} and the baseline of \textsf{MixInt}.
On the contrary, \textsf{MmixInt} and \textsf{MixInt} baseline achieve the better prediction accuracies than the \textsf{Disc} baseline.
Comparing \textsf{MmixInt} and the \textsf{MixInt} baseline, the former one achieves the lower transmission latency and power consumption.
This is due to the fact that in \textsf{MmixInt}, the input dimension of the first convolutional layer in camera~B (having a lower frame rate) is smaller.
This means less multiplications at camera~B and smaller upload payload size for forward propagation signals, leading to lower power consumption and transmission latency.

Subsequently, we compare different methods for aggregating image features in terms of prediction accuracy and power consumption in BS in Fig.~\ref{fig:comprehensive_comparison}(c).
In terms of prediction accuracy, \textsf{MmixAgg} exhibits slightly poorer performance than \textsf{ConcAgg}.
This accuracy loss can be explained by the reduced model fitting capability due to the input dimension reduction and weight parameter reductions.
Meanwhile, the accuracy loss can be of the order of 1\% in every method for balancing a frame rate difference, hence, the accuracy loss is not severe.
In addition, proposed \textsf{MmixAgg} could reduce the power consumption at the BS as shown in Figs~\ref{fig:comprehensive_comparison}(c).
This is due to the fact that \textsf{MmixAgg} results in fewer input dimension of the fully connected layer (see. Fig.~\ref{fig:compare_hetslagg_hetslfedavg}(a)).
This input dimension reduction leads to the fewer numbers of weight parameters and the reduction of the energy cost for the memory accesses to load the parameters.
These results provide insights into the fact that with the proposed \textsf{MmixAgg}, one can feasibly reduce the energy cost at the BS without severely sacrificing the prediction accuracy relative to the  \textsf{ConcAgg} baseline.

\section{Conclusions}
\label{sec:conclusions}
We proposed a novel mmWave received power prediction framework that fuses both RF signals and heterogeneous visual data in a communication-and-energy-efficient manner while preserving privacy.
To this end, focusing on preserving privacy first, we proposed HetSLAgg that splits the NN models into several segments and distribute the lower segments into the cameras.
The upper segment held in the BS combines the image feature activations uploaded from cameras with the RF feature activations and performs predictions based on the combined feature activations.
Subsequently, we addressed the issue of achieving better tradeoff communication-and-energy-efficiency and prediction accuracy owing to the usage of multiple visual data.
Specifically, we proposed BS-side manifold mixup-based interpolation to make visual data with different frame rates compatible with one another while achieving better tradeoff between  communication-and-energy-efficiency and prediction accuracy.
We also proposed an energy-scalable feature aggregation methods so that the input dimension of the upper segments of the NN layer in the BS does not scale up, thereby avoiding the increase in power consumption at the BS.
The experimental results demonstrated that the designed HetSLAgg enhances prediction accuracy by leveraging heterogeneous modalities while achieving better tradeoff between communication energy-efficiency and prediction accuracy.

Although our ideas can be applied to settings of more than two distributed cameras, the key concern is that the usage of more cameras does not necessarily lead to better performances due to the heterogeneity of both data qualities and channel qualities.
For example, in terms of data qualities, some cameras may not have visual data containing informative features for recognizing pedestrian movement owing to limited FoV, which do contribute to prediction accuracy enhancement.
In terms of channel qualities, some cameras could have lower channel gains, which could delay the overall training performances.
Thus, how to detect and schedule informative cameras among the deployed ones to balance prediction accuracies and communication efficiencies could be an interesting topic for our future work.

\bibliographystyle{IEEEtran}
\bibliography{IEEEabrv,main}

\end{document}